\documentclass[a4paper,11pt]{article}

\usepackage{jcappub} 

\usepackage[T1]{fontenc} 
\usepackage{subfigure}

\title{The evolution of the FRW universe with decaying metastable dark energy --- a dynamical system analysis}

\author[a,b,1]{Marek Szyd{\l}owski,\note{Corresponding author.}}
\author[a]{Aleksander Stachowski,}
\author[c]{Krzysztof Urbanowski}

\affiliation[a]{Astronomical Observatory, Jagiellonian University, \\ Orla 171, 30-244 Krak{\'o}w, Poland}
\affiliation[b]{Mark Kac Complex Systems Research Centre, Jagiellonian University, \\ {\L}ojasiewicza 11, 30-348 Krak{\'o}w, Poland}
\affiliation[c]{Institute of Physics, University of Zielona G{\'o}ra, \\ Prof. Z. Szafrana 4a, 65-516 Zielona G{\'o}ra, Poland}

\emailAdd{marek.szydlowski@uj.edu.pl}
\emailAdd{aleksanderms@wp.pl}
\emailAdd{K.Urbanowski@if.uz.zgora.pl}

\abstract{
We investigate a cosmological model in which dark energy identified with the vacuum energy which is running and decaying. In this model vacuum is metastable and decays into a bare (true) vacuum. This decaying process has a quantum nature and is described by tools of the quantum decay theory of unstable systems. We have found formulas for an asymptotic behavior of the energy density of dark energy in the form of a series of inverse powers of the cosmological time. We investigate the dynamics of FRW models using dynamical system methods as well as searching for exact solutions. From dynamical analysis we obtain different evolutional scenarios admissible for all initial conditions. For the interpretation of the dynamical evolution caused by the decay of the quantum vacuum we study the thermodynamics of the apparent horizon of the model as well as the evolution of the temperature. For the early Universe, we found that the quantum effects modified the evolution of the temperature of the Universe. In our model the adiabatic approximation is valid and the quantum vacuum decay occurs with an adequate unknown particle which constitutes quantum vacuum. We argue that the late-time evolution of metastable energy is the holographic dark energy.}
\arxivnumber{1812.00616}

\begin{document}
\maketitle
\flushbottom

\section{Introduction}

While the standard cosmological model successfully describes astronomical observation from the primordial nucleosynthesis to the present day, this effective description contains many troubles. In the present cosmology, in principle, we meet two following problems. The first one is the cosmological constant problem. It is a unexplained difference between the measured value of the vacuum energy and the value calculated using quantum field theory methods \cite{Weinberg:1988cp}. The second problem is the need of explanation why the present density of matter and dark energy in the Universe has the same order of magnitude \cite{Weinberg:2000yb}. This question is called the coincidence problem.

Results of the current astronomical observations lead to the conclusion that the Universe is in an accelerated phase \cite{Ade:2015rim}. This acceleration is explained as a consequence of a presence of dark energy. The analysis of astronomical observations shows that there is a tension between local and primordial measurements of cosmological parameters \cite{Ade:2015rim}. A possible explanation of this tension can be dark energy which changes in time \cite{DiValentino:2017rcr}. In this paper, we consider dark energy dependent on time, $\rho_{\text{de}} = \rho_{\text{de}}(t)$ and it is metastable. We assume that dark energy decays with the increasing time $t$ to $\rho_{\text{bare}}$ ($\rho_{\text{de}}(t) \to \rho_{\text{bare}} \neq 0$ when $t \to \infty$). The decaying vacuum energy was considered in many papers (see e.g. \cite{Krauss:2007rx,Krauss:2008pt} and also \cite{Urbanowski:2011zz,Urbanowski:2016pks}). Recently models with metastable dark energy have drawn attention in the context of discrepancies appearing in the standard cosmological model related with the so-called $H_0$ tension problem \cite{Li:2019san,Pan:2019hac}.

Shafieloo et al. \cite{Shafieloo:2016bpk} assumed that $\rho_{\text{de}}(t)$ decays according to the radioactive exponential decay law. But, this assumption is not sufficient to explain the evolution of the Universe in the time of the decaying process of dark energy, because the creation of the Universe is a quantum process. So, the metastable dark energy was propose as the value of a scalar field at the false vacuum state and the decay of the dark energy should be considered as the quantum decay process.

The quantum decay processes consist of the following phases \cite{Fonda:1978dk,Peshkin:2017elo}:
\begin{itemize}
\item the early time initial phase,
\item the canonical or exponential phase,
\item the late time non-exponential phase.
\end{itemize}
In result, the first phase and the third one are missed in the case of the radioactive decay law only. The theoretical analysis of quantum decaying processes shows that for the late time, the survival probability of the system, which is considered in its initial state (i.e. the decay law), should tend to zero as $t \to \infty$ much more slowly than an exponential function of time and that as a function of time it has an inverse power-like form at this regime of time \cite{Fonda:1978dk,Khalfin:1957cdt}. (This last effect was confirmed experimentally by Rothe et al. in 2006 \cite{Rothe:2006}).
The consequences of the decay process of the dark energy as the quantum decay process can be found only if to use a quantum decay law to describe decaying metastable dark energy. In our analysis of this problem, we use the idea presented by Krauss and Dent \cite{Krauss:2007rx,Krauss:2008pt} and our research is a direct continuation of the idea presented therein and its development initiated in \cite{Urbanowski:2011zz,Urbanowski:2016pks}. It was also studied by Szydlowski et al.~\cite{Szydlowski:2017wlv}.
This idea results from the observation that the general form of the quantum decay law, the properties of the survival amplitude and thus the properties of the energy of the system in a unstable state do not depend on the form of the Hamiltonian (or Lagrangian) containing interactions causing the quantum decay process \cite{Krylov:1947tmi,Fock:1978fqm,Khalfin:1957cdt,Fonda:1978dk}. According to this idea in order to find some general properties of the system in the meta-stable false vacuum state it is sufficient to express the corresponding survival probability in the form of the Fourier transform of the energy density distribution function $\omega (E)$ (see more detailed discussion, e.g. in \cite{Krauss:2007rx}). (The function $\omega(E)$ is the probability to find the energy of the system in the unstable state between energies $E$ and $E + dE$).
So the advantage of this approach is that the conclusions are general and do not depend on the choice of the form of Hamiltonian (or Lagrangian). Only the values of the decay rate $\Gamma_{0}$ and the energy $E_{0}$ the system in the unstable state considered measured at the exponential phase depend on the interactions (that is on the Hamiltonian) and this information as well as the information about the decay products is contained in the form of the energy density distribution function $\omega (E)$. It is because a meta-stable state is not an eigenvector of the Hamiltonian of the system considered. In result such a meta-stable state can be expanded in the complete base of the energy eigenvectors of the Hamiltonian and simply $\omega (E)$ equals to the square of the modulus of the expansion coefficients. This property is the basis of the so-called Fock-Krylov theory of unstable states later used by Khalfin and others (see, e.g. \cite{Krylov:1947tmi,Fock:1978fqm,Khalfin:1957cdt,Fonda:1978dk}) and this is why $\omega (E)$ contains also information about the decay products and other quantities characterizing the system. So, generally it is enough to assume that the quantum vacuum decays in order to find the general properties of this process and this observation is used in our paper.

In cosmology, the Hawking temperature and entropy with the apparent horizon can be considered in an analogous way to as it is considered in the context of the black hole horizon \cite{Cai:2005ra, Gibbons:1977mu}. For cosmological models, the apparent horizon always exists even if the horizon does not exist. We study the problem of thermodynamics in cosmological models with an interaction between matter and dark energy.

The paper has the following structure. Section~2 contains preliminaries. The standard approach in metastable (false) vacuum studies is analyzed and briefly discussed in this section. In section~3, there is an introduction into a formalism of a quantum decaying process of dark energy. Section~4 is about time scales in the process of decaying metastable dark energy. In section~5, we investigate cosmology with a decaying dark energy. Thermodynamics of the cosmological model with an interaction between matter and dark energy is considered in section~6. For a deeper interpretation of decaying process we consider an evolution of temperature in section~7. Section~8 contains conclusions.

\section{Preliminaries}

\begin{figure}[t]
\begin{center}
\includegraphics[width=70mm]{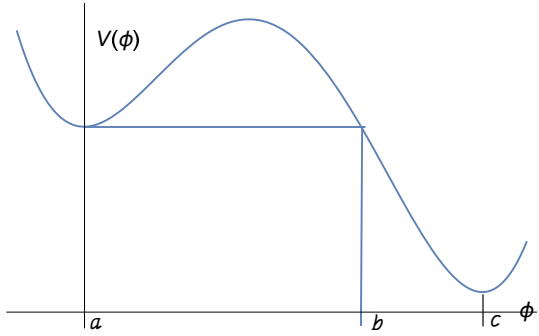}
\caption{Example of the potential $V(\phi)$ having a local minimum $V(a)$ for $\phi = a$, which corresponds to the state $|0\rangle^{\text{F}}$ and the true minimum $V(c)$ for $\phi = c$ corresponding to the true vacuum state $|0\rangle^{\text{T}}$.}
\label{h1}
\end{center}
\end{figure}

Here we show that starting points of the standard approach in metastable (false) vacuum studies and of our approach described in the next section are the same. We analyze in this section consequences of the standard approach (see, e.g., seminal papers \cite{Coleman:1977py,Callan:1977pt}) and a consequences following from this approach for models of decaying dark energy. Coleman et al.~\cite{Coleman:1977py,Callan:1977pt} discussed the instability of a physical system, which is not at an absolute energy minimum, and which is separated from the absolute minimum by an effective potential barrier. They showed that if the early Universe is too cold to activate the energy transition to the minimum energy state then a quantum decay, from the false vacuum to the true vacuum, is still possible through a barrier penetration via the macroscopic quantum tunneling. To find how a particle evolves from the point corresponding to the value $a = \phi(\vec{r}_{a},t_{a})$ of the scalar field $\phi(\vec{r},t)$ for which the potential $V(\phi)$ has a local minimum (see figure~\ref{h1}) to end at the points corresponding to the values of $\phi(\vec{r},t)$ for $|\vec{r}| > |\vec{r}_{b}|$ and times $t> t_{b}$, where $\vec{r}$ is a vector describing the position, and $t$ is the time one should calculate the amplitude describing this process. (The potential $V(\phi)$ reaches the absolute (true) minimum for $\phi = c = \phi(\vec{r}_{c},t_{c})$, where $|\vec{r}_{c}| > |\vec{r}_{b}| > |\vec{r}_{a}|$ and $t_{c} > t_{b} > t_{a}$). The states vectors corresponding to these cases are denoted as follows: $|0\rangle^{\text{F}}$ denotes the state of the analyzed system being in the local minimum of $V(\phi)$ and the vector $|0\rangle^{\text{T}}$ is the state of the system corresponding to absolute (true) minimum of $V(\phi)$.
Putting for simplicity $t_{a}=0$ we can express the amplitude describing the evolution of the particle from the point corresponding to $\phi =a$ to points corresponding the position $|\vec{r}| > |\vec{r}_{b}|$ and times $t > t_{b}$ as follows

\begin{equation}
{\cal Z}_{av}(t) =\,^{\text{F}}\langle 0|e^{\textstyle{-\frac{i}{\hbar}\mathfrak{H}t}}|0\rangle^{\text{V}}. \label{C0}
\end{equation}
where $\mathfrak{H}$ is the self-adjoint Hamiltonian, $ t>t_{b}$, $v =a$ or $v=c$ and $\text{V}$ denotes $\text{F}$ (false vacuum) or $\text{T}$ (true vacuum).
The mathematical tool used within the standard approach to calculate amplitudes of the type (\ref{C0}) is Feynman's path integral method.
the amplitude ${\cal Z}_{av}(t)$
can be expressed as follows:
\begin{equation}
{\cal Z}_{av}(t) \equiv N \int e^{\textstyle{iS[\phi(x(t))]}}\,D[x], \label{C1}
\end{equation}
where $N$ is the normalization constant, $S[\phi(x(t))]$ is the action, $D[x]$ is the path element.
Within this approach the energy of the system in the false vacuum state $|0\rangle^{\text{F}}$ is extracted from the diagonal matrix element ${\cal Z}_{aa}(t) = \,^{\text{F}}\langle 0|e^{\textstyle{-\frac{i}{\hbar}\mathfrak{H}t}}|0\rangle^{\text{F}}$ calculated for $t > t_{b} $, which is called the \emph{persistence amplitude} (see, e.g., \cite{Andreassen:2016cff}) and it is identified with the \emph{survival amplitude} in the quantum theory of unstable states. The state $|0\rangle^{\text{F}}$ is not an eigenstate of $\mathfrak{H}$ and it can be expanded using the basis formed from the normalized eigenstates of the energy operator $\mathfrak{H}$,
\begin{equation}
\mathfrak{H} |E_{n}\rangle = E_{n}|E_{n} \rangle, \label{E-n}
\end{equation}
which gives
\begin{equation}
{\cal Z}_{aa}(t) \equiv \,^{\text{F}}\langle 0|e^{\textstyle{-\frac{i}{\hbar}\mathfrak{H} t}}|0\rangle^{\text{F}} = \sum_{n}e^{\textstyle{-\frac{i}{\hbar}E_{n} t}} | \langle E_{n}|0 \rangle^{\text{F}}|^{2}, \label{C2}
\end{equation}
where within this approach $E_{0}$ denotes the minimal energy of the system. Next, the time is rotated to the complex axis $t \to T=it$. Then ${\cal Z}_{aa}(t)\,\to\, Z_{aa}(T)$, and
\begin{equation}
Z_{aa}(T) = \,^{\text{F}}\langle 0|e^{\textstyle{-\frac{T}{\hbar}\mathfrak{H}}}|0\rangle^{\text{F}} \, =\,
\sum_{n}\,e^{\textstyle{-\frac{T}{\hbar} E_{n}}} | \langle E_{n}|0 \rangle^{\text{F}}|^{2}, \label{C2+T}
\end{equation}
or,
\begin{equation}
Z_{aa}(T) \equiv
N \int e^{\textstyle{- S[\phi(x(t))]}}\,D[x]. \label{C3}
\end{equation}
Now by letting $T \to \infty$ all the states with energy higher than $E_{0}$ are exponentially suppressed in comparison to the term involving $|E_{0}\rangle$,
\begin{equation}
\sum_{n}\,e^{\textstyle{-\frac{T}{\hbar} E_{n}}} | \langle E_{n}|0 \rangle^{\text{F}}|^{2}
\underset{T \to \, \infty}{\thicksim} | \langle E_{0}|0 \rangle^{\text{F}}|^{2} \, e^{\textstyle{ -\,E_{0} \frac{T}{\hbar}}}, \label{C3a}
\end{equation}
Of course such an assumption cancels possible early time ($t > t_{b} > t_{a} =0$) and asymptotically late time effects appearing in the amplitude ${\cal Z}_{aa}(t)$. So for $T \gg T_{b}$, where $T_{b} = it_{b}$, that is for $T\to \infty$ in (\ref{C2+T}) it is sufficient to take into account only the the dominating element for $T \to \infty$ in (\ref{C2+T}), and then one obtains (see \cite{Callan:1977pt})
\begin{equation}
Z_{aa}(T) = \,^{\text{F}}\langle 0|e^{\textstyle{-\frac{T}{\hbar}\mathfrak{H}}}|0\rangle^{\text{F}} \simeq e^{\textstyle{-\frac{T}{\hbar}E_{0}}}\, |\langle E{_0}|0 \rangle^{\text{F}}|^{2}.
\label{C3b}
\end{equation}
More accurate treatment of the problem (see, e.g. \cite{Patrascioiu:1981vv}) shows that in the area of long times, the the expression (\ref{C3b}) for the amplitude $Z_{aa}(T)$ is valid only for finite time and for $T \to \infty$ the exponential term (\ref{C3b}) in $Z_{aa}(T)$ is completed by terms proportional to powers of $\frac{1}{T}$.
Indeed, as i has been shown that (see formula (1.4) and also formulae (2.8), (2.11), (3.11) in \cite{Patrascioiu:1981vv})
\begin{equation}
Z_{aa}(T) \underset{T \to \, \infty}{\thicksim} |b|^{2}\left[e^{\textstyle{-(E_{0} - i\frac{\Gamma}{2})\frac{T}{\hbar}}}\,-\, d\left( \frac{\Gamma}{E_{0}}\right)^{\frac{5}{2}} \left(\frac{\hbar}{\Gamma T} \right)^{\frac{3}{2}} \right], \label{pet2-8}
\end{equation}
where $b$ and $d$ are constant (see \cite{Patrascioiu:1981vv}) and $\Gamma$ is the decay rate. According to conclusion drawn in \cite{Patrascioiu:1981vv} this result implies that only for $t = -iT$ such that,
\begin{equation}
\left( \frac{\Gamma t}{\hbar} \right)^{\frac{3}{2}} e^{\textstyle{-\frac{\Gamma t}{\hbar}}} \,\gg\, \left(\frac{\Gamma}{E_{0}}\right)^{\frac{5}{2}},
\label{pet-2-8a}
\end{equation}
the exponential decay law is reasonable accurate. When for later times the expression on the left of the inequality (\ref{pet-2-8a}) becomes smaller than the right the decay law take the form of powers of $1/t$. An analogous result can be found in \cite{Wipf:1985mr} (see formula (1.2) therein) or in \cite{Andreassen:2016cvx} (see formulae (A.13), (A.18)--(A.20) in \cite{Andreassen:2016cvx}). Unfortunately, this problem is unknown to the vast majority of researchers using and developing the standard approach initiated by Coleman's work, in which eq.~(\ref{C3b}) is the basis for further calculations.

The conclusion resulting from eq.~(\ref{C3b}) is that
\begin{equation}
\frac{E_{0}}{\hbar}\,T \simeq \ln\,[ |\langle E{_0}|0 \rangle^{\text{F}}|^{2}] \,-\,\ln\, \left[\,^{\text{F}}\langle 0|e^{\textstyle{-\frac{T}{\hbar}\mathfrak{H}}}|0\rangle^{\text{F}}\right], \label{C4}
\end{equation}
and this relation is used to find the energy of the system in the false vacuum state $|0\rangle^{\text{F}}$,
\begin{equation}
E_{0}^{\text{F}}\simeq \,-\,\lim_{T\to \infty}\,\left[\frac{\hbar}{T}\,\ln[\,^{\text{F}}\langle 0|e^{\textstyle{-\frac{T}{\hbar}\mathfrak{H}}} |0\rangle^{\text{F}}]\right] =
\,-\,\lim_{T\to \infty}\,\left(\frac{\hbar}{T}\,\ln\,[Z_{aa}(T)]\right),
\label{C5}
\end{equation}
where within the standard approach $Z_{aa}(T)$ is calculated using eq.~(\ref{C3}). From eq.~(\ref{C4}) it follows that for very large $T \to \infty$ the real part $\Re\,[E_{0}^{\text{F}}]$ of $E_{0}^{\text{F}}$ given by eq.~(\ref{C5}) can be interpreted as the energy of the system $E^{\text{F}}$ in the false vacuum state plus some corrections: $E^{\text{F}} = \Re\, E_{0}^{\text{F}} \simeq E_{0} + ({\rm corrections})$. Within the standard approach discussed in this section eq.~(\ref{C5}) is used to define the decay rate $\Gamma_{F}$ as the imaginary part of $E_{0}^{\text{F}}$,
\begin{equation}
\Gamma_{F} \simeq\,\,2 \Im\,\left(\lim_{T\to \infty}\, \frac{\hbar}{T}\,\ln[Z_{aa}(T)] \right),
\label{C6}
\end{equation}
and $Z_{aa}(T)$ is calculated using eq.~(\ref{C3}).
Equations (\ref{C1})--(\ref{C6}) are the essence of the standard approach used to study cosmological models with meta-stable false vacuum or decaying dark energy and running $\Lambda$ but taking into account conclusion resulting from eqs~(\ref{pet2-8}), (\ref{pet-2-8a}) one can be afraid that eq.~(\ref{C6}) may lead to the wrong result.

Note that starting from eq.~(\ref{C3b}) we can find $E_{0}$ equivalently by differentiating two sides of eq.~(\ref{C3b}) with respect to $T$ and then dividing two sides of such obtained a new equation by eq.~(\ref{C3b}) respectively. This leads to the following relation,
\begin{equation}
E_{0} \simeq \mathfrak{E} \equiv \mathfrak{E}(T) \stackrel{\rm def}{=} - \frac{\hbar}{Z_{aa}(T)}\, \frac{\partial Z_{aa}(T)}{\partial T}. \label{E(T)-1}
\end{equation}
 Within the method discussed one assumes that $\mathfrak{E}(T) \neq ( \mathfrak{E}(T) )^{\ast}$ and $\Gamma_{F} = 2\Im\,[\mathfrak{E}(T)]$, and for $T > T_{b}$ but $T\ll \infty$ one obtains the same decay rate as in the case of the radioactive decay of the dark energy.

The relation (\ref{E(T)-1}) can be rewritten using real times $t$ instead of Euclidean $T=it$: Then $Z_{aa}(T) \to {\cal Z}_{aa}(t)$ and $\mathfrak{E}(T) \to
\epsilon (t)$,
\begin{equation}
\mathfrak{E}(T) \to \epsilon (t) \stackrel{\rm def}{=} \frac{i\hbar}{{\cal Z}_{aa}(t)}\, \frac{\partial {\cal Z}_{aa}(t)}{\partial t}, \label{e(t)-1}
\end{equation}
and here $\epsilon (t) = E_{\text{F}}(t) - \frac{i}{2}\Gamma_{\text{F}}(t)$. For $t > t_{b}$ but $t \ll \infty$ one obtains that $\Gamma_{\text{F}}(t) \simeq \Gamma_{\text{F}}$.

Now let us compare results obtained for $Z_{aa}(T)$ given by (\ref{pet2-8}) using eq.~(\ref{C6}) and (\ref{e(t)-1}). From (\ref{pet2-8}) it follows for the regime of the exponential decay, that is when the condition (\ref{pet-2-8a}) holds, that
\begin{equation}
\Gamma_{F} \simeq\,\,2 \Im\,\left(\lim_{T\to \infty}\, \frac{\hbar}{T}\,\ln[Z_{aa}(T)] \right) \equiv \Gamma, \label{GF=G}
\end{equation}
whereas for later times, when the exponential term in (\ref{pet2-8}) becomes negligible comparing the term proportional to $(\frac{\hbar}{T})^{\frac{3}{2}}$, the result following from (\ref{C6}) is $\Gamma_{F} =0$. On the other hand if one starts from eq. (\ref{e(t)-1}) taking $t = - iT$ and using ${\cal Z}(t)$ then one finds for the exponential decays regime that $\Gamma_{F} = \Gamma$, but for the later times one obtains that $\Gamma_{F} = \Gamma_{F}(t) \underset{T \to \, \infty}{\thicksim} \frac{3 \hbar}{t} \equiv 3 \hbar\,\exp[- \ln\,t]$.

At the end of the discussion of the standard approach in studies of meta-stable vacuum we should stress it that from mathematical point of view amplitudes ${\cal Z}_{aa}(t)$ defined in eq.~(\ref{C0}) (or in eq.~(\ref{C2})) and the energy $\epsilon (t)$ defined by the relation (\ref{e(t)-1}) are identical to quantities corresponding to them and used in the next section of this paper, i.e. to the survival amplitude ${\cal A}(t)$ and the effective Hamiltonians $h(t)$ governing the time evolution in a one-dimensional subspace found in the quantum theory of unstable states. This similarity becomes quite obvious if we take into account that quantum tunneling can be used to model quantum decay processes (see, e.g., \cite{Winter:1961zz,Calzetta:2006rg,vanDijk:1999zz,vanDijk:2002ru,Dicus:2002tdq,Martorell:2009qpd,vanDijk:2019}).

\section{Decay of a dark energy as a quantum decay process}

When one studies quantum unstable systems, the survival probability (the decay law)
\begin{equation}
{\cal P}(t) = |{\cal A}(t)|^{2}, \label{P-t}
\end{equation}
is used to describe changes in time of an unstable system. This means that properties properties of the survival amplitudes
\begin{equation}
{\cal A}(t) = \langle \psi| \psi(t)\rangle
\label{amp}
\end{equation}
are analyzed in such a case. Here a vector $|\psi\rangle$ represents the unstable state of the system considered and $| \psi (t)\rangle $ is the solution of the Schr\"{o}dinger equation
\begin{equation}
i \hbar \frac{\partial}{\partial t} |\psi(t)\rangle = \mathfrak{H}
|\psi (t)\rangle. \label{Sch}
\end{equation}
The initial condition for eq.~(\ref{Sch}) in the case considered is usually assumed to be
\begin{equation}
| \psi (t = t_{0} \equiv 0) \rangle \stackrel{\rm def}{=}
| \psi\rangle, \quad \text{or equivalently}, \quad {\cal A}(0) = 1. \label{A(0)}
\end{equation}
In eq.~(\ref{Sch}) $\mathfrak{H}$ denotes the complete (full), self-adjoint Hamiltonian of the system.

Using the basis in ${\cal H}$ build from normalized eigenvectors $|E\rangle,\,\ E\in \sigma_{c}(\mathfrak{H}) = [E_{\text{min}}, {\infty})$ of $\mathfrak{H}$ and using the expansion of $|\psi\rangle$ in this basis one can express the amplitude ${\cal A}(t)$ as the following Fourier integral
\begin{equation}
{\cal A}(t) \equiv {\cal A}(t - t_{0}) = \int_{E_{\text{min}}}^{\infty} \omega(E)\,
e^{\textstyle{-\,\frac{i}{\hbar}\,E\,(t - t_{0})}}\,d{E},
\label{a-spec}
\end{equation}
where $\omega(E) = \omega(E)^{\ast}$ and $\omega(E) > 0$ is the probability to find the energy of the system in the state $|\psi\rangle$ between $E$ and $E\,+\,dE$ and $E_{\text{min}}$ is the minimal energy of the system. The last relation (\ref{a-spec}) means that the survival amplitude ${\cal A}(t)$ is a Fourier transform of an absolute integrable function $\omega (E)$. If we apply the Riemann-Lebesgue lemma to the integral (\ref{a-spec}) then one concludes that there must be ${\cal A}(t) \to 0$ as $t \to \infty$. This property and the relation (\ref{a-spec}) are an essence of the Fock-Krylov theory of unstable states \cite{Krylov:1947tmi,Fock:1978fqm}.

So, within this approach the amplitude ${\cal A}(t)$, and thus the decay law ${\cal P}(t)$ of the unstable state $|\psi\rangle$, are determined completely by the density of the energy distribution $\omega(E)$ for the system in this state \cite{Krylov:1947tmi,Fock:1978fqm} (see also \cite{Fonda:1978dk,Kelkar:2010qn}, and so on. (This approach is also applicable to models in quantum field theory~\cite{Giacosa:2011xa,Giacosa:2018dzm}).

As it was mentioned in the previous section, it is assumed that the transition from meta-stable false vacuum to the true vacuum can occur through quantum tunneling. Within the quantum theory such transitions through quantum tunneling can be used to model a decay process of an unstable state. What is more this process can be described using the Fock-Krylov theory (see e.g. \cite{Winter:1961zz,vanDijk:1999zz,vanDijk:2002ru,Dicus:2002tdq,Martorell:2009qpd,vanDijk:2019}). Strictly speaking in the case of the quantum tunneling used to model the quantum decay process the survival probability can be also expressed in the form of the Fourier transform as it was done in eq.~(\ref{a-spec}). This means that the general formalism based on the Fock-Krylov theory is fully suitable for describing the properties of a decaying false vacuum and a running dark energy.

Note that in fact the amplitude ${\cal A}(t)$ contains information about the decay law ${\cal P}(t)$ of the state $|\psi\rangle$, that is about the decay rate $\gamma_{\psi}$ of this state, as well as the energy ${E}_{\psi}$ of the system in this state. This information can be extracted from ${\cal A}(t)$. It can be done rigorously using the equation governing the time evolution in the subspace of unstable states, ${\cal H}_{\parallel} \ni |\psi\rangle_{\parallel} \equiv |\psi \rangle$. Such an equation follows from the Schr\"{o}dinger equation (\ref{Sch}) for the total state space ${\cal H}$.

Using Schr\"{o}dinger equation (\ref{Sch}) one finds that within the problem considered
\begin{equation}
i \hbar \frac{\partial}{\partial t}\langle\psi |\psi (t) \rangle = \langle \psi|\mathfrak{H} |\psi (t)\rangle. \label{h||1}
\end{equation}
From this relation one can conclude that the amplitude $A(t)$ satisfies the following equation
\begin{equation}
i \hbar \frac{\partial {\cal A}(t)}{\partial t} = h(t)\,{\cal A}(t), \label{h||2}
\end{equation}
where
\begin{equation}
h(t) = \frac{\langle \psi|\mathfrak{H} |\psi (t)\rangle}{{\cal A}(t)} \equiv \frac{i\hbar}{{\cal A}(t)}\,\frac{\partial{\cal A}(t)}{\partial t}, \label{h(t)}
\end{equation}

The effective Hamiltonian $h(t)$ governs the time evolution in the subspace of unstable states ${\cal H}_{\parallel}= \mathbb{P} {\cal H}$, where $\mathbb{P} = |\psi\rangle \langle \psi|$ (see \cite{Urbanowski:1994epq} and also \cite{Urbanowski:2006mw,Urbanowski:2008kra,Urbanowski:2009lpe} and references therein). The subspace ${\cal H} \ominus {\cal H}_{\parallel} = {\cal H}_{\perp} \equiv \mathbb{Q} {\cal H}$ is the subspace of decay products. Here $\mathbb{Q} = \mathbb{I} - \mathbb{P}$. One meets the effective Hamiltonian $h(t)$ when one starts with the Schr\"{o}dinger equation for the total state space ${\cal H}$ and looks for the rigorous evolution equation for a distinguished subspace of states ${\cal H}_{||} \subset {\cal H}$ \cite{Urbanowski:1994epq,Giraldi:2015ldu,Giraldi:2016zom}. In general $h(t)$ is a complex function of time and in the case of ${\cal H}_{\parallel}$ of two or more dimension the effective Hamiltonian governing the time evolution in such a subspace is a non-hermitian matrix $H_{\parallel}$ or non-hermitian operator. There is
\begin{equation}
h(t) = E_{\psi}(t) - \frac{i}{2} {\gamma}_{\psi}(t), \label{h-m+g}
\end{equation}
where
$
E_{\psi}(t) = \Re\,[h(t)]$, ${\gamma}_{\psi}(t) = -2\,\Im\,[h(t)],
$
are the instantaneous energy (mass) $E_{\psi}(t)$ and the instantaneous decay rate, ${\gamma}_{\psi}(t)$. (Here $\Re\,(z)$ and $\Im\,(z)$ denote the real and imaginary parts of $z$, respectively).

The quantity ${\gamma}_{\psi}(t) = -2\,\Im\,[h(t)]$ is interpreted as the decay rate, because it satisfies the definition of the decay rate used in quantum theory. Simply, using (\ref{h(t)}) it is easy to check that
\begin{equation} \label{G-equiv}
\frac{{\gamma}_{\psi}(t)}{\hbar} \stackrel{\rm def}{=}
- \frac{1}{{\cal P}(t)} \frac{\partial {\cal P}(t)}{\partial t} =
- \frac{1}{|{\cal A}(t)|^{2}}\,\frac{\partial |{\cal A}(t)|^{2}}{\partial t}
\equiv - \frac{2}{\hbar}\,\Im\,[h(t)].
\end{equation}
As it is seen from the above definition of the decay rate, $\gamma_{\psi}(t)$ is precisely defined according to the conditions formulated in \cite{Andreassen:2016cvx,Kristiano:2018oyv}, where the precision decay rate calculations in quantum field theory are presented. On the other hand eq.~(\ref{G-equiv}) shows that the definition of $\gamma_{\psi}(t)$ as the imaginary part of the instantaneous Hamiltonian $h(t)$ (or $\epsilon (t)$ --- see eq.~(\ref{e(t)-1})), $\gamma_{\psi}(t) = - 2\Im\,[h(t)] \equiv - 2\Im\,[\epsilon(t)]$, is the same as that called the ``precision definition'' in \cite{Andreassen:2016cvx,Kristiano:2018oyv} and used therein to analyze properties of the the decay rate of the metastable false vacuum.

We have $| \psi (t) \rangle = \exp\,[-\frac{i}{\hbar}t\mathfrak{H}] | \psi\rangle$. So, in a general case ${\cal A}(t) \equiv $ \linebreak $ \langle \psi|\exp\,[-\frac{i}{\hbar}t\mathfrak{H}] | \psi\rangle$. It is not difficult to see that this property and hermiticity of $\mathfrak{H}$ imply that \cite{Fonda:1978dk}
\begin{equation}
({\cal A}(t))^{\ast} ={\cal A}(-t). \label{amp-ast}
\end{equation}

The conclusion resulting from this property and from the relation (\ref{h(t)}) is that
\begin{equation}
h(-t) = \left(h(t)\right)^{\ast}. \label{h=h-ast}
\end{equation}
Therefore there must be
\begin{equation}
E_{\psi}(-t) = E_{\psi}(t) \quad \text{and} \quad \gamma_{\psi}(-t) = - \gamma_{\psi}(t), \label{even+odd}
\end{equation}
That is, the instantaneous energy
$ E_{\psi}(t) = \Re\,[h(t)]$ is an even function of time $t$ and the instantaneous decay rate $ {\gamma}_{\psi}(t) = -2\,\Im\,[h(t)]$ an odd function of $t$.

In the extensive literature many quantum unstable systems are described within the Fock-Krylov theory using the Breit-Wigner energy density distribution function $\omega_{BW}(E)$. The use of $\omega_{BW}(E)$ is convenient because it describes relatively well a large class of unstable systems and allows to find an analytical form of the survival amplitude $a(t)$ (see, eg. \cite{Sluis:1991dqs} and papers cited therein). It turns out that the decay curves obtained in this simplest case are very similar in form to the curves calculated for the more general $\omega (E)$, (see \cite{Kelkar:2010qn} and the analysis in \cite{Fonda:1978dk}). What is more, it appears that the decay of the false-vacuum state by quantum tunneling can be described to a good approximation using the energy density distribution function having the Breit-Wigner form (see, e.g. \cite{Calzetta:2006rg}). Generally $\omega_{BW}(E)$ is a good approximation when one uses the Fock-Krylov approach to describe the tunneling as the model of the quantum decay process. So, to find the most typical properties of the decay process it is sufficient to make the relevant calculations for $\omega (E)$ modeled by the Breit-Wigner distribution of the energy density $\omega_{BW}(E)$. For such $\omega (E) = \omega_{BW}(E)$ one can find relatively easy an analytical form of $a(t)$ at very late times as well as an analytical asymptotic form of $h(t)$, $E(t)$ and $\gamma (t)$ for such times.
\begin{equation}
\omega (E) \equiv \omega_{\text{BW}}(E) \stackrel{\text{def}}{=} \frac{N}{2\pi} \,
\frac{\Theta (E - E_{\text{min}}){\Gamma}_{0}}{(E-E_{0})^{2} +
(\frac{{\Gamma}_{0}}{2})^{2}}, \label{omega-bw}
\end{equation}
where $N$ is a normalization constant and $\Theta (E) $ is the unit step function.

The parameters $E_{0}$ and ${\Gamma}_{0}$ correspond to the energy of the system in the unstable state $|\psi\rangle$ and its decay rate at the exponential (or canonical) regime of the decay process. $E_{\text{min}}$ is the minimal (the lowest) energy of the system. In Section 1 it was said that $\omega (E)$ contains information characterizing the given unstable state: In the case $\omega (E) = \omega_{BW}(E)$ quantities $E_{0}$, ${\Gamma}_{0}$ and $E_{\text{min}}$ are exactly the parameters characterizing the unstable state considered. The different values of these parameters correspond to different unstable states.

Inserting $\omega_{\text{BW}}(E)$ into formula (\ref{a-spec}) for the amplitude ${\cal A}(t)$ and assuming for simplicity that $t_{0} = 0$, after some algebra one finds that
\begin{equation}
{\cal A}(t) = \frac{N}{2\pi}\,
e^{\textstyle{- \frac{i}{\hbar} E_{0}t }}\,{\cal I}_{\beta}\left(\frac{{\Gamma}_{0} t}{\hbar}\right) \label{I(t)a}
\end{equation}
where
\begin{equation}
{\cal I}_{\beta}(\tau) \stackrel{\rm def}{=}\int_{-\beta}^{\infty}
 \frac{1}{\eta^{2} + \frac{1}{4}}\, e^{\textstyle{- i\eta\tau}}\,d\eta. \label{I(t)}
\end{equation}
Here $\tau = \frac{{\Gamma}_{0}\,t}{\hbar} \equiv \frac{t}{\tau_{0}}$, $\tau_{0}$ is the lifetime, $\tau_{0} = \frac{\hbar}{{\Gamma}_{0}}$, and $\beta = \frac{E_{0} - E_\text{min}}{{\Gamma}_{0}} > 0$. The integral ${\cal I}_{\beta}(\tau)$ has the following structure
\begin{equation}
{\cal I}_{\beta}(\tau) = {\cal I}_{\beta}^{\text{pole}}(\tau) + {\cal I}_{\beta}^{L}(\tau) \label{Ip+IL}
\end{equation}
where
\begin{equation}
 {\cal I}_{\beta}^{\text{pole}}(\tau) = \int_{- \infty}^{\infty}
 \frac{1}{\eta^{2}
+ \frac{1}{4}}\, e^{\textstyle{ -i \eta \tau}} \, d\eta
\equiv 2 \pi\,e^{\textstyle{-\,\frac{\tau}{2}}} \label{Ip}
\end{equation}
 and
\begin{equation}
 {\cal I}_{\beta}^{L}(\tau) = - \int_{+ \beta}^{\infty}
 \frac{1}{\eta^{2} + \frac{1}{4}} \,
 e^{\textstyle{ + i \eta \tau}} \, d\eta. \label{IL}
\end{equation}
The integral ${\cal I}_{\beta}^{L}(\tau)$ can be expressed in terms of an integral-exponential function \cite{Sluis:1991dqs,Urbanowski:2006mw,Urbanowski:2009lpe,Raczynska:2018ofh} (for a definition, see \cite{Olver:2010hmf,Abramowitz:1964hmf}). The result (\ref{Ip+IL}) means that there is a natural decomposition of the survival amplitude ${\cal A}(t)$ into two parts
\begin{equation}
{\cal A}(t) = {\cal A}_{c}(t) + {\cal A}_{L}(t), \label{Ac+AL}
\end{equation}
where
\begin{equation}
{\cal A}_{c}(t) = \frac{N}{2\pi}\,
e^{\textstyle{ - \frac{i}{\hbar} E_{0}t }}\,{\cal I}_{\beta}^{\text{pole}}\left(\frac{{\Gamma}_{0} t}{\hbar}\right) \equiv N\,e^{\textstyle{ - \frac{i}{\hbar} E_{0}t }}\,
e^{\textstyle{- \frac{{\Gamma}_{0}\,t}{2\hbar}}}, \label{Ac(t)}
\end{equation}
and
\begin{equation}
{\cal A}_{L}(t) = \frac{N}{2\pi}\,
e^{\textstyle{ - \frac{i}{\hbar} E_{0}t }}\,{\cal I}_{\beta}^{L}\left(\frac{{\Gamma}_{0} t}{\hbar}\right), \label{AL(t)}
\end{equation}
${\cal A}_{c}(t)$ is the canonical part of the amplitude ${\cal A}(t)$ describing the pole contribution into ${\cal A}(t)$ and ${\cal A}_{L}(t)$ represents the remaining part of ${\cal A}(t)$.

From the decomposition (\ref{Ac+AL}) it follows that in the general case within the model considered the survival probability (\ref{P-t}) contains the following parts
\begin{equation} \label{Ac+AL-2}
\begin{split}
{\cal P}(t) &= |{\cal A}(t)|^{2} \equiv |{\cal A}_{c}(t) + {\cal A}_{L}(t)|^{2} \\
&= |{\cal A}_{c}(t)|^{2} \,+ \,2\,\Re\,[{\cal A}_{c}(t)\,( {\cal A}_{L}(t) )^{\ast}]\,+\,|{\cal A}_{L}(t)|^{2}.
\end{split}
\end{equation}
This last relation is especially useful when one looks for a contribution of a late time properties of the quantum unstable system into the survival amplitude.

The late time form of the integral ${\cal I}_{\beta}^{L}(\tau)$ and thus the late time form of the amplitude ${\cal A}_{L}(t)$ can be relatively easy to find using analytical expression for ${\cal A}_{L}(t)$ in terms of the integral-exponential functions or simply performing the integration by parts in (\ref{IL}). One finds for $t \to \infty$ (or $\tau \to \infty$) that the leading term of the late time asymptotic expansion of the integral ${\cal I}_{\beta}^{L}(\tau)$ has the following form for $\tau \to \infty$ \cite{Raczynska:2018ofh}
\begin{equation}
\begin{split}
{\cal I}_{\beta}^{L}(\tau) &\simeq \frac{i}{\tau} \frac{e^{\textstyle{i\beta \tau}}}{\beta^{2} + \frac{1}{4}}
\Big[-1 + \frac{2 \beta}{\beta^{2} + \frac{1}{4}} \frac{i}{\tau} + \frac{2}{\beta^{2} + \frac{1}{4}}\Big( 1- \frac{4 \beta^{2}}{\beta^{2} + \frac{1}{4}}\Big) \Big(\frac{i}{\tau}\Big)^{2} \\
&+\frac{24\beta}{(\beta^{2} + \frac{1}{4})^{2}}\Big(\frac{2\beta^{2}}{\beta^{2} + \frac{1}{4}} - 1 \Big)\,\Big(\frac{i}{\tau}\Big)^{3} \\
&+\frac{24}{(\beta^{2} + \frac{1}{4})^{2}}\Big(-\frac{16\beta^{4}}{(\beta^{2} + \frac{1}{4})^{2}}+\frac{12\beta^{2}}{\beta^{2} + \frac{1}{4}}-1\Big)
\Big(\frac{i}{\tau}\Big)^{4} + \cdots \Big], \label{IL-as}
\end{split}
\end{equation}
Thus inserting (\ref{IL-as}) into (\ref{AL(t)}) one can find late time form of ${\cal A}_{L}(t)$, which is in agreement with the general result obtained in \cite{Urbanowski:2008kra}.
Note that according to the remark after eq. (\ref{omega-bw}) the parameter $\beta$ is build from parameters characterizing the given unstable state and thus this information is contained in all asymptotic expansions of the type (\ref{IL-as}) or similar one.

Now let us analyze properties of the instantaneous energy $E(t)$ and instantaneous decay rate $\gamma_{\psi} (t)$ in the model considered. These quantities are defined using the effective Hamiltonian $h(t)$. In order to find $h(t)$ we need for the quantity $i\,\hbar\, \frac{\partial {\cal A}(t)}{\partial t}$ (see (\ref{h(t)})). From eq.~(\ref{I(t)a}) one finds that
\begin{equation}
i \hbar \frac{\partial {\cal A}(t)}{\partial t} = E_{0} \,{\cal A}(t)
+ \Gamma_{0}\,\frac{N}{2\pi}\,e^{\textstyle{-\frac{i}{\hbar} E_{0}t}}\,{\cal J}_{\beta}(\tau(t)), \label{J-R-1}
\end{equation}
where
\begin{equation}
{\cal J}_{\beta}(\tau) = \int_{- \beta}^{\infty}\,\frac{x}{x^{2} + \frac{1}{4}}\,e^{\textstyle{-ix\tau}}\,dx, \label{J-R}
\end{equation}
or simply (see (\ref{I(t)})),
\begin{equation}
{\cal J}_{\beta}(\tau) \equiv i\frac{\partial {\cal I}_{\beta} (\tau)}{\partial \tau}. \label{J-R-eq}
\end{equation}
This last relation (\ref{J-R-eq}) is a very convenient way to obtain an analytical expression for ${\cal J}_{\beta}(\tau) $ after finding an analytical formula for ${\cal I}_{\beta} (\tau)$.

Now the use of (\ref{I(t)a}), (\ref{J-R-1}) and (\ref{h(t)}) leads to the conclusion that within the model considered,
\begin{equation}
h(t) = i \hbar \frac{1}{{\cal A}(t)}\,\frac{\partial {\cal A}(t)}{\partial t} = E_{0} + \Gamma_{0}\,\frac{{\cal J}_{\beta}(\tau(t))}{{\cal I}_{\beta}(\tau(t))}, \label{h(t)-1}
\end{equation}
which means that
\begin{equation}
E(t) = \Re\,[h(t)] = E_{0} + \Gamma_{0}\,\Re\,\left[\frac{{\cal J}_{\beta}(\tau(t))}{{\cal I}_{\beta}(\tau(t))}\right], \label{E(t)-2}
\end{equation}
and
\begin{equation}
\gamma_{\psi}(t) = -2\,\Im[h(t)] = -2\,\Gamma_{0}\,\Im\left[\frac{{\cal J}_{\beta}(\tau(t))}{{\cal I}_{\beta}(\tau(t))}\right]. \label{g(t)-a}
\end{equation}

In order to visualize properties of $E(t)$ it is convenient to use the following function
\begin{equation}
\kappa (t) \stackrel{\rm def}{=} \frac{E(t) - E_{\text{min}}}{E_{0} - E_{\text{min}}} . \label{kappa}
\end{equation}
Using (\ref{E(t)-2}) one finds that
\begin{equation}
E(t) - E_{\text{min}} = E_{0} - E_{\text{min}} + \Gamma_{0}\,\Re\,\Big[\frac{{\cal J}_{\beta}(\tau)}{{\cal I}_{\beta}(\tau)}\Big], \label{E(t)-1}
\end{equation}
If we divide two sides of equation (\ref{E(t)-1}) by $E_{0} - E_{\text{min}}$ then one obtains the function $\kappa (t)$ (see (\ref{kappa})) we are looking for
\begin{equation}
\kappa (\tau(t)) = 1 + \frac{1}{\beta}\,\Re\,\Big[\frac{{\cal J}_{\beta}(\tau(t))}{{\cal I}_{\beta}(\tau(t))}\Big]. \label{kappa-1}
\end{equation}

In order to find the asymptotic form of $h(t)$ one needs the late time form for ${\cal J}_{\beta}(\tau)$. Proceeding in the same way as in the case of ${\cal I}_{\beta}^{L}(\tau)$ one finds for $\tau \to \infty$ \cite{Raczynska:2018ofh} that
\begin{equation}
\begin{split}
{\cal J}_{\beta}(\tau) &\simeq \frac{i}{\tau}\,\frac{e^{\textstyle{i\beta \tau}}}{\beta^{2} + \frac{1}{4}}
\Big\{\beta +
\Big[1 - \frac{2 \beta^{2}}{\beta^{2} + \frac{1}{4}}\Big]\,\frac{i}{\tau} + \frac{2\beta}{\beta^{2} + \frac{1}{4}}\Big[\frac{4 \beta^{2}}{\beta^{2} + \frac{1}{4}} - 3\Big]\,\Big(\frac{i}{\tau}\Big)^{2}\\
&+ \frac{6}{\beta^{2} + \frac{1}{4}}\Big[-\frac{8\beta^{4}}{(\beta^{2} + \frac{1}{4})^{2}} + \frac{8\beta^{2}}{\beta^{2} + \frac{1}{4}}-1 \Big]\,\Big(\frac{i}{\tau}\Big)^{3} \\
&+ \frac{24\beta}{(\beta^{2} + \frac{1}{4})^{2}}\Big[ \frac{16\beta^{4}}{(\beta^{2} + \frac{1}{4})^{2}} -\frac{20\beta^{2}}{\beta^{2} + \frac{1}{4}}+5 \Big]\,\Big(\frac{i}{\tau}\Big)^{4}\,
\ldots \Big\}. \label{J-as}
\end{split}
\end{equation}
Starting from the asymptotic expression (\ref{J-as}) and using formula (\ref{h(t)-1}) one can find the late time asymptotic form of $h(t)$ and thus of $E(t)$ and $\gamma_{\psi} (t)$ for the model considered \cite{Raczynska:2018ofh}
\begin{equation} \label{Re-h-as}
\begin{split}
{E(t)\, \vline}_{\,t \rightarrow \infty} &= {\Re\,[h(t)] \, \vline}_{t \to \infty}
\simeq { E}_{\text{min}}\, -\,2\,
\frac{ { E}_{0}\,-\,{E}_{\text{min}}}{ \Gamma_{0}^{2}\,(\beta^{2} + \frac{1}{4}) } \,
\left(\frac{\hbar}{t} \right)^{2} + \\
&+\,\frac{1}{4}\,\frac{6 - 21 \beta + 48\beta^{2} - 64 \beta^{3} - 288 \beta^{4} + 464\beta^{5}}{\Gamma_{0}^{3}\,(\beta^{2} + \frac{1}{4})^{4}}\,\left(\frac{\hbar}{t} \right)^{4} + \cdots,
\end{split}
\end{equation}
and,
\begin{equation} \label{Im-h-as}
{{\it\gamma}_{\psi}(t)\, \vline}_{\,t \rightarrow \infty} = - 2 \Im\,[h(t)]
\simeq 2\,\frac{\hbar}{t} +\frac{1}{2}\, \frac{1 + 24 \beta - 28 \beta^{2} - 96 \beta^{3} + 64\beta^{4}}{\Gamma_{0}^{2}\,(\beta^{2} + \frac{1}{4})^{3}}\,
\left(\frac{\hbar}{t}\right)^{3} + \cdots \, .
\end{equation}

In the general case in the agreement with properties (\ref{even+odd}) we have for $t \to \infty$
\begin{equation}
{E(t)\vline}_{\,t \rightarrow \infty} \simeq{ E}_{\text{min}} +
\sum_{k \geq 1 } f_{2k}\,(\frac{\hbar}{t})^{2k}, \label{E-gen}
\end{equation}
i.e., $ {E(t)\vline}_{\,t \rightarrow \infty}$ is an even function of time $t$, and
\begin{equation}
{{\it\gamma}_{\psi}(t)\vline}_{\,t \rightarrow \infty} \simeq
\sum_{k \geq 0} g_{2k+1}\, (\frac{\hbar}{t})^{2k+1}. \label{gamma-gen}
\end{equation}
is an odd function of $t$, where $f_{2k}= (f_{2k})^{\ast}$ and $g_{2k+1}=(g_{2k+1})^{\ast}$.

All the above described properties of the quantum unstable system are general one and do not depend on a specific form of interactions forcing the decay process or a mechanism responsible for such a process. Simply, in all quantum decay processes general properties of the decay law are the same. The decay law has always an initial non-exponential phase, the exponential (canonical) phase and the late time non-exponential phase. This concerns also such a quantum process as the decaying dark energy.

\section{Different time scales in the process of decaying metastable dark energy}

In our approach the parameterization of metastable dark energy is calculated from the quantum mechanics principles and incorporated into the FRW cosmology. Then we consider the Friedmann equation with this quantum correction. The physical processes take place in different time scales. For our approximation it is important to satisfy the adiabatic approximation: time scale of quantum processes is much less than the cosmological time scale and proportional to $1/H$ where $H$ is the Hubble parameter. In this section we present different time scales related to processes of decay of metastable dark energy for demonstration of validation of the adiabatic approximation.

The question arises when the quantum properties of evolution become negligibly small and cease to have a significant impact on the further evolution of the system. The observation that the ability to interfere is an essential and inherent feature of quantum processes suggests an answer for this question. As one can see our formula (\ref{Ac+AL-2}) contains the interference element. Namely, the element $2\,\Re\,[{\cal A}_{c}(t)\,( {\cal A}_{L}(t) )^{\ast}]$ in the formula (\ref{Ac+AL-2}) for survival probability ${\cal P}(t)$ describes the contribution to the survival probability coming from the interference of the pole contribution ${\cal A}_{c}(t)$ and the ${\cal A}_{L}(t)$, which dominates at late times, to the full survival probability ${\cal P}(t)$. Hence the reasonable conclusion is that for such late times $t$, for which the following inequality will be fulfilled,
\begin{equation}
2\left|\,\Re\,[{\cal A}_{c}(t)\,( {\cal A}_{L}(t) )^{\ast}]\,\right|\, \ll \,\left|{\cal A}_{L}(t)\right|^{2}, \label{inter1}
\end{equation}
the quantum nature of temporal evolution will become negligible.

Now let us consider as an example the model based on the Breit-Wigner energy distribution function $\omega_{BW}(E)$. Within this model ${\cal A}_{c}(t)$ is given by the formula (\ref{Ac(t)}). The late form of the amplitude ${\cal A}_{L}(t)$ can be obtained by inserting the asymptotic expansion (\ref{IL-as}) into (\ref{AL(t)}). So we obtain for $t \to \infty$,
\begin{equation}
{\cal A}_{L}(t) = - \frac{N}{2\pi}\,e^{\textstyle{ - \frac{i}{\hbar} \Sigma_{0}\tau }}\,\frac{i}{\tau}\,\frac{e^{\textstyle{i \beta \tau}}}{\beta^{2} + \frac{1}{4}} + \cdots,
\label{AL(t)-2}
\end{equation}
where $\Sigma_{0} = \frac{E_{0}}{\Gamma_{0}}$. Hence, at late times
\begin{equation}
2\,\Re\,[{\cal A}_{c}(\tau)\,({\cal A}_{L}(\tau))^{\ast}] \simeq \frac{N^{2}}{2 \pi}\,\frac{1}{\beta^{2} + \frac{1}{4}}\,\frac{\sin\,\beta \tau}{\tau}\,e^{\textstyle{-\frac{\tau}{2}}}, \label{inter2}
\end{equation}
and
\begin{equation}
\left|{\cal A}_{L}(\tau)\right|^{2} \simeq \frac{N^{2}}{4 \pi^{2}}\,\frac{1}{(\beta^{2} + \frac{1}{4})^{2}}\,\frac{1}{\tau^{2}}. \label{|AL|}
\end{equation}

The form of $2\,\Re\,[{\cal A}_{c}(\tau)\,({\cal A}_{L}(\tau))^{\ast}]$ and $\left|{\cal A}_{L}(\tau)\right|^{2}$ as functions of time $\tau = \frac{\Gamma_{0} t}{\hbar}$ for different values of $\beta$ is presented below in figures~\ref{f1-all} and \ref{f2-all}.
\begin{figure}[ht]
\centering
\subfigure[\scriptsize{The case $\beta = 1$.}]{
\includegraphics[width=60mm]{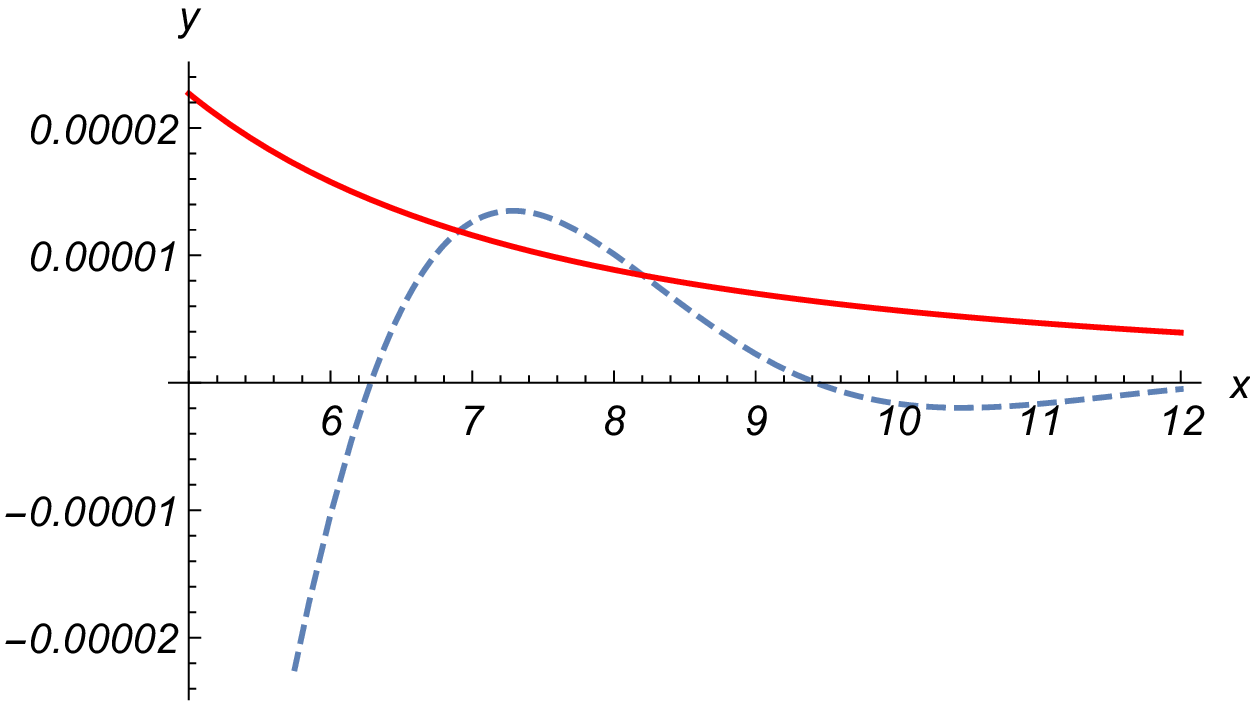}
\label{f1}
}
\qquad
\subfigure[\scriptsize{The case $\beta= 1$.}]{
\includegraphics[width=60mm]{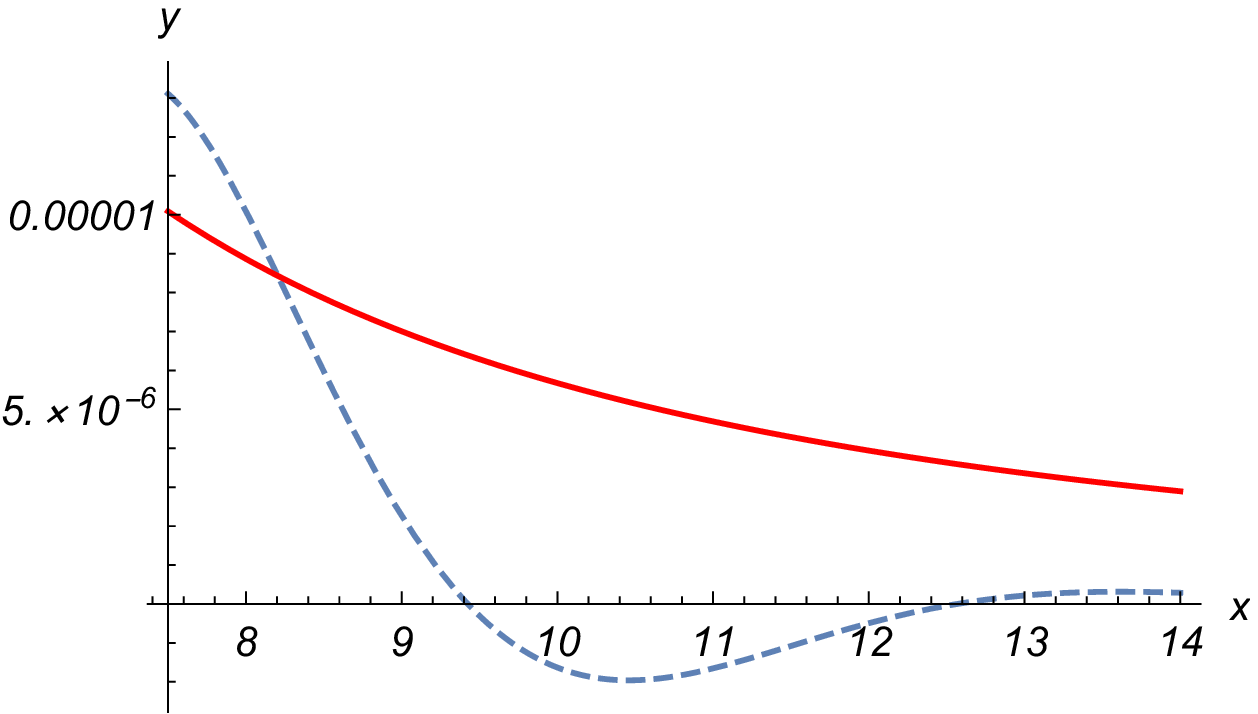}
\label{f2}
} \\

\subfigure[\scriptsize{The case $\beta = 10$.}]{
\includegraphics[width=60mm]{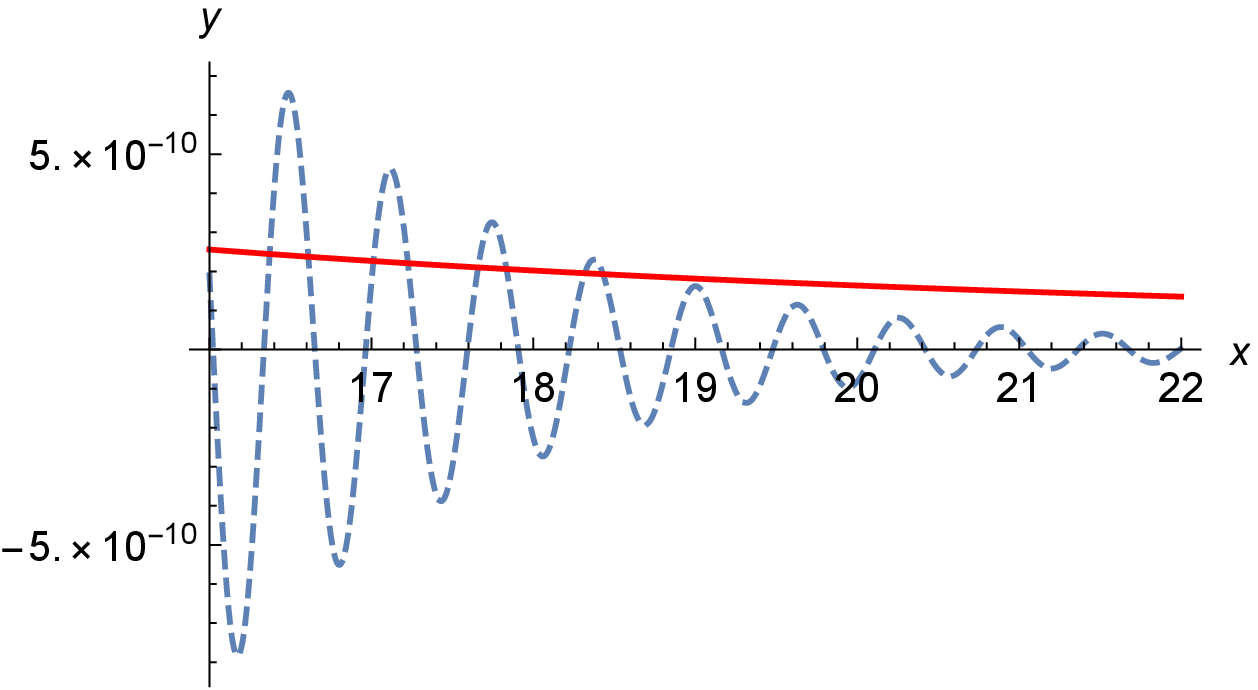}
\label{f3}
}
\qquad
\subfigure[\scriptsize{The case $\beta= 100$.}]{
\includegraphics[width=60mm]{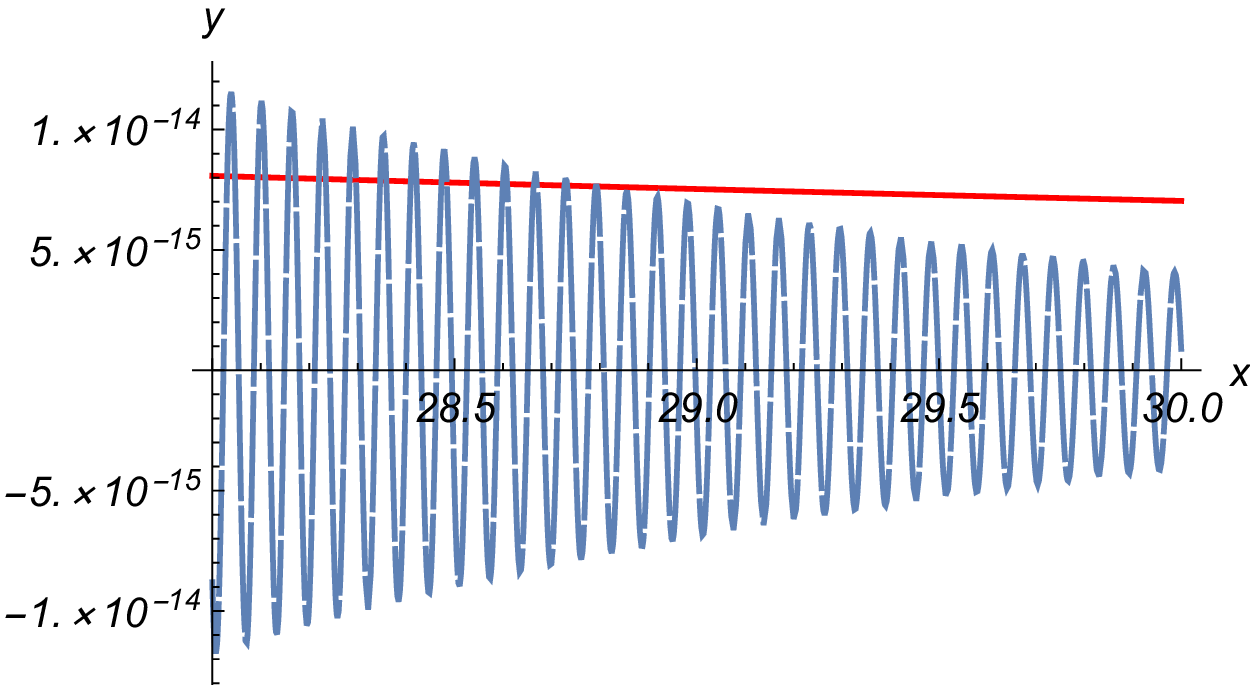}
\label{f4}
}
\caption{Contributions of $2\,\Re\,[{\cal A}_{c}(\tau)\,({\cal A}_{L}(\tau))^{\ast}]$ and $\left|{\cal A}_{L}(\tau)\right|^{2}$ to the survival probability ${\cal P}(t)$ as functions of time $t$ at late times.
Axes: $y$ = $2\,\Re\,[{\cal A}_{c}(\tau)\,({\cal A}_{L}(\tau))^{\ast}]$ (Dashed line), $\left|{\cal A}_{L}(\tau)\right|^{2}$ (Solid line); $x = \frac{\Gamma_{0} t}{\hbar}$.}
\label{f1-all}
\end{figure}

\begin{figure}[ht]
\centering
\subfigure[\scriptsize{The case $\beta = 10$.}]{
\includegraphics[width=60mm]{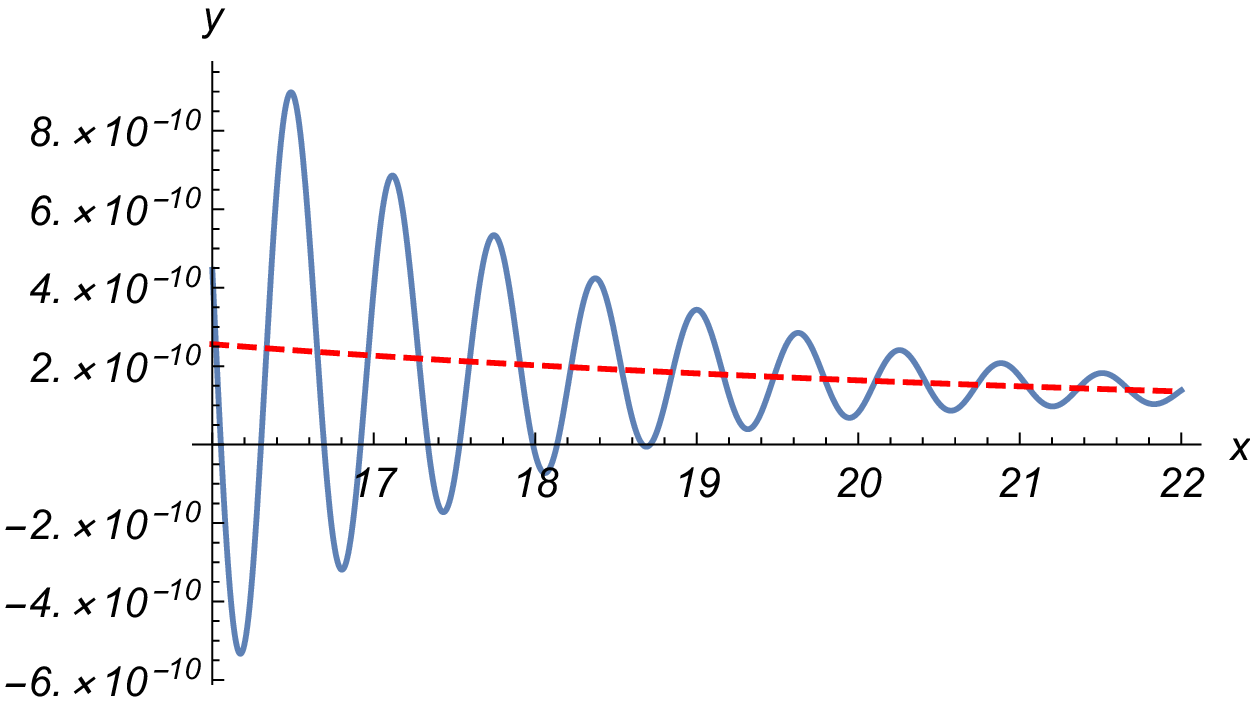}
\label{f4a}
}
\qquad
\subfigure[\scriptsize{The case $\beta= 10$.}]{
\includegraphics[width=60mm]{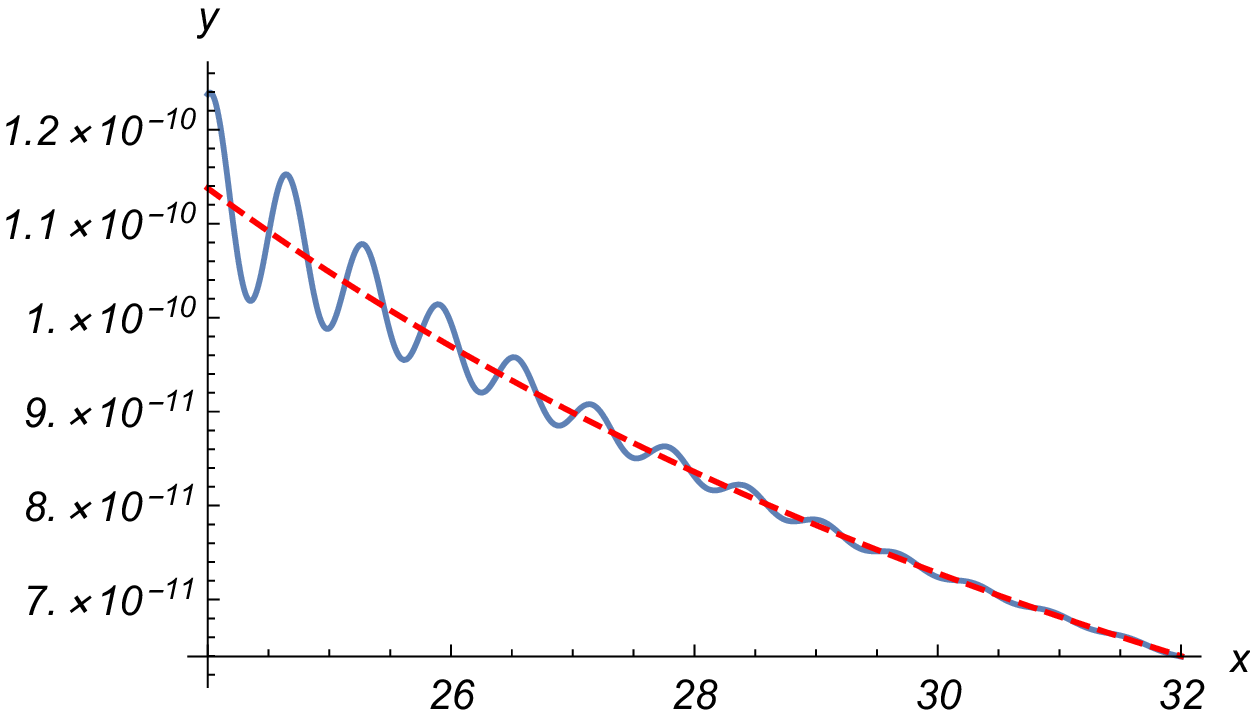}
\label{f4b}
} \\

\subfigure[\scriptsize{The case $\beta = 10$.}]{
\includegraphics[width=60mm]{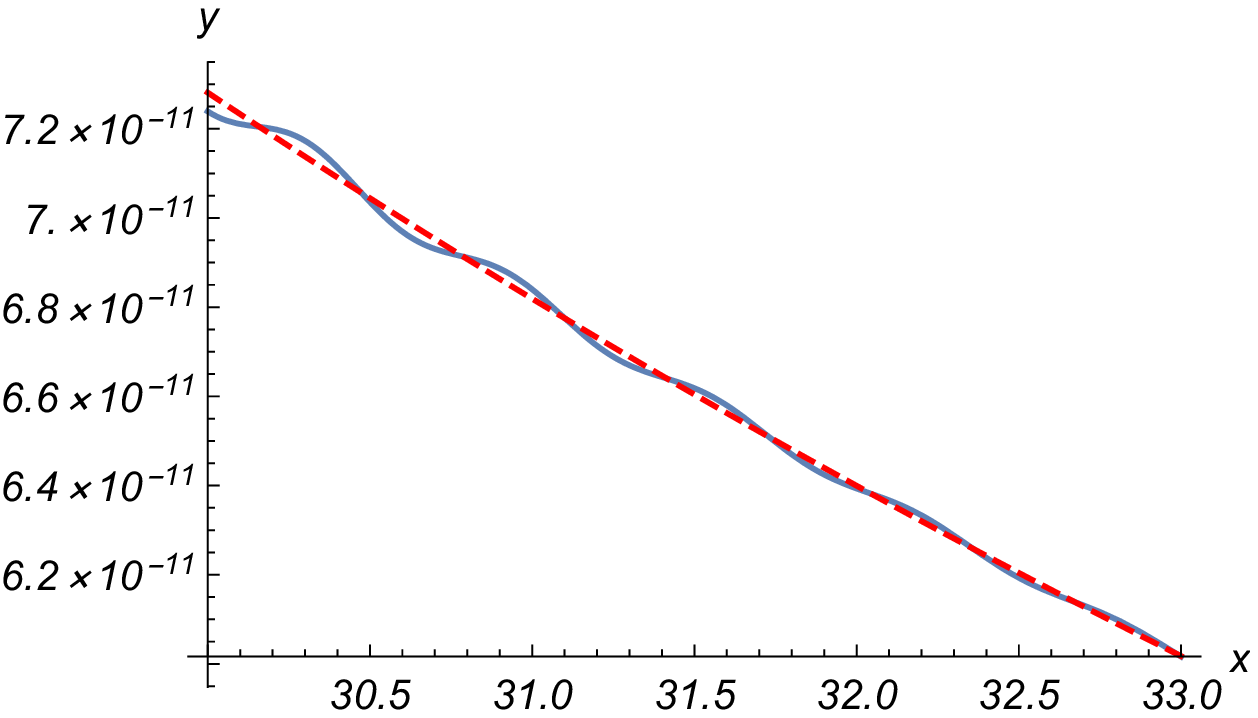}
\label{f4c}
}
\caption{A comparison of contributions of the sum $2\,\Re\,[{\cal A}_{c}(\tau)\,({\cal A}_{L}(\tau))^{\ast}]$ and $\left|{\cal A}_{L}(\tau)\right|^{2}$ with the contribution of $\left|{\cal A}_{L}(\tau)\right|^{2}$ to the survival probability ${\cal P}(t)$ as functions of time $t$ at late times. Axes: $y$ = $2\,\Re\,[{\cal A}_{c}(\tau)\,({\cal A}_{L}(\tau))^{\ast}] + \left|{\cal A}_{L}(\tau)\right|^{2}$ (Solid line), $\left|{\cal A}_{L}(\tau)\right|^{2}$ (Dashed line); $x = \frac{\Gamma_{0} t}{\hbar}$.}
\label{f2-all}
\end{figure}

\begin{figure}[t]
\centering
\includegraphics[width=70mm]{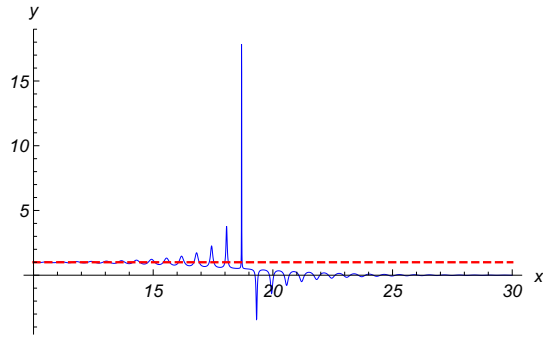}
\caption{An illustration of the typical behavior of energy $E(t)$ over a period of time containing the instant $t=T_\text{q-c}$. The case $\beta = 10$. Axes:
$y =\kappa (t) \equiv \frac{E(t) - E_\text{min}}{E_{0} - E_\text{min}}$ --- the solid line; $\kappa (t)$ for $E(t) = E_{0} = \text{const}$. --- the dashed line; $x = \frac{\Gamma_{0} t}{\hbar}$.}
\label{f-E}
\end{figure}

Within the model considered we obtain from eq.~(\ref{inter2}) that
\begin{equation}
2\left|\,\Re\,[{\cal A}_{c}(t)\,( {\cal A}_{L}(t) )^{\ast}]\,\right| \leq \frac{N^{2}}{2 \pi}\,\frac{1}{\beta^{2} + \frac{1}{4}}\,\frac{1}{\tau}\,e^{\textstyle{-\frac{\tau}{2}}}, \label{inter3}
\end{equation}
and the right hand side of the inequality (\ref{inter3}) is the upper bound for the values of $2\,\Re\,[{\cal A}_{c}(t)\,( {\cal A}_{L}(t) )^{\ast}]$. This means that in our model in order to find an approximate value of time $T_\text{q-c}$ from which the quantum effects will be negligibly small, it is sufficient to use the simplified version of the condition (\ref{inter1}). It can be obtained replacing the left hand side of the inequality (\ref{inter1}) by its upper bound (\ref{inter3}) and the right hand side of (\ref{inter1}) by (\ref{|AL|}) respectively. This leads to the following condition
\begin{equation}
e^{\textstyle{- \frac{\tau}{2}}} \ll \frac{1}{2\pi}\,\frac{1}{\beta^{2} + \frac{1}{4}}\,\frac{1}{\tau}. \label{inter4}
\end{equation}
The time $T_\text{q-c}$, we are looking for, is the solution of the following equation
\begin{equation}
e^{\textstyle{- \frac{\tau}{2}}} = \frac{1}{2\pi}\,\frac{1}{\beta^{2} + \frac{1}{4}}\,\frac{1}{\tau}, \label{Tqc1}
\end{equation}
and according to the condition (\ref{inter1}) for times $t \gg T_\text{q-c}$ the quantum nature of the time evolution of the system considered will be negligible. (Here $T_\text{q-c}$ means $T_\text{quantum-to-classical}$). There are the following solutions of eq.~(\ref{Tqc1}) in the cases presented in figures~\ref{f1}, \ref{f2}, \ref{f3} and \ref{f4}. We have $\tau_\text{q-c} = 8.37177$ for $\beta =1$, $\tau_\text{q-c} = 18.7539$ for $\beta =10$ and $\tau_\text{q-c} = 28.8185$ for $\beta =100$, (where $\tau_\text{q-c} = \frac{\Gamma_{0} \,T_\text{q-c}}{\hbar}$). So, the large ratio $\frac{E_{0} - E_\text{min}}{\Gamma_{0}}$, (that is $\beta$), the large $\tau_\text{q-c}$, i.e., $T_\text{q-c}$. Analyzing results presented in figures~\ref{f1-all} and \ref{f2-all} one can see that starting from times $\tau \simeq \tau_\text{q-c}$ the contribution of the interference term $2\Re\,[{\cal A}_{c}(t)\,( {\cal A}_{L}(t) )^{\ast}]$ into the survival probability becomes negligible small in comparison to the contribution of $|{\cal A}_{L}(t)|^{2}$ and at these times simply ${\cal P}(t) \simeq |{\cal A}_{L}(t)|^{2}$, which can be interpreted that for times $t > T_\text{q-c}$ the quantum nature of time evolution practically disappears. This last conclusion is illustrated in a graphical form in figure~\ref{f2-all}, where one can see that within the model considered the contribution of the sum, $2\,\Re\,[{\cal A}_{c}(\tau)\,({\cal A}_{L}(\tau))^{\ast}]\,+\,\left|{\cal A}_{L}(\tau)\right|^{2}$, to the survival probability ${\cal P}(t)$ for $t > T_\text{q-c}$ together with the rise of time $t$ becomes closer and closer to the contribution coming only from $\left|{\cal A}_{L}(\tau)\right|^{2}$ so that ${\cal P}(t) \simeq |{\cal A}_{L}(t)|^{2}$ for $t \gg T_\text{q-c}$. The another conclusion is that this analysis highlights the fact that relations (\ref{Re-h-as}) and (\ref{Im-h-as}) are valid only for $t \gg T_\text{q-c}$. In other words, the energy $E(t)$ for $t \gg T_\text{q-c}$ has the form (\ref{Re-h-as}) to a very good approximation, while there is $E(t) \sim E_{0}$ for $t \ll T_\text{q-c}$ (see figure~\ref{f-E}). So the following quantum effect takes place: there is $\left. E(t) \right|_{t\to \infty} \ll E_{0}$ for $t \gg T_\text{q-c}$ and at this time region the energy $E(t) \simeq \left. E(t) \right|_{t\to \infty}$ can be approximated as the sum of the minimal energy $E_\text{min}$ of the system and time-depended corrections, which leading element is of order $1/t^{2}$ (see eq.~(\ref{Re-h-as}) and figure~\ref{f-E}).

Now, how does the above picture look in the case of real elementary particles? Let us consider as an example the Higgs boson. Its mass is $E_{0} = m_{H} \simeq 125.2$ GeV, the decay width $\Gamma_{0} = \Gamma_{H} \simeq 4.2$ MeV. The lifetime $\tau_{0} = \tau_{H} = \frac{\hbar}{\Gamma_{H}}$ is $\tau_{H} \simeq 1.57 \times 10^{-22}$ s. Assuming for simplicity that $E_{\text{min}} =0$ we obtain that $\beta = \frac{E_{0} - E_{\text{min}}}{\Gamma_{H}} \equiv \frac{m_{H}}{\Gamma_{H}} \stackrel{\rm def}{=} \beta_{mx} \simeq 29809.5$. Inserting this maximal value $\beta_{mx}$ of $\beta$ into eq.~(\ref{Tqc1}) we obtain $\tau_\text{q-c} \simeq 52.82$. This means that within the model defined by the Breit-Wigner function of the energy distribution the time $T_\text{q-c}$ equals: $T_\text{q-c} = \tau_\text{q-c} \times \tau_{H} \simeq 8.3 \times 10^{-21}$ s. This value corresponds to the epoch of radiation domination. Hence we can conclude that considering the quantum decay processes of types similar to the Higgs boson decays one can expect that at times much later than $t = T_\text{q-c} \simeq 8.3 \times 10^{-21}$ s the quantum nature of the time evolution will no longer have a significant impact on the evolution of the system considered.

Applying quantum theory to describe the process of a false vacuum decay, or more generally, to describe a decaying dark energy, we meet a certain incoherence caused by the fact that equations of the evolution of the Universe (Einstein's equations) and thus their solutions are symmetric with respect to transformation: $t\,\rightarrow\,(-t)$, what can not be said about quantum decay processes. In general the time reversal symmetry should be valid in the quantum system until the initial conditions are taken into account. In quantum mechanics, if one wants to examine the evolution in a given state of the system in time, one must first prepare the initial state of this system. The process of preparing this state obviously breaks the symmetry of reflection in time, and it is mathematically expressed in a concrete choice of initial conditions. In other words the process of preparing the system in a given quantum unstable state breaks the time reversal symmetry (see, eg. \cite{Urbanowski:2000ud}). Simply, one prepares (creates) the state of the system at a set initial instant of time $t=t_{0}$, which then evolves in time $t>t_{0}$. If this state was created at the initial moment of time $t_{0}$, it means that in earlier moments of time $t < t_{0}$ it was not (or in other words, that it was not occupied), and in turn, it means that the process of the preparation of the system in this state broke time reversal symmetry.

Analyzing the problem of the time reversal symmetry in cosmology one should invoke a problem of baryogenesis and the baryon-antibaryon asymmetry. From the astrophysics observations we know that the observed Universe is composed almost entirely of matter with little or no primordial antimatter: There is the observed imbalance in baryonic matter and antibaryonic matter in the observable Universe. Sakharov \cite{Sakharov:1967rr} has formulated conditions that ensure that a small baryon asymmetry may have been produced in quantum processes occurring in the early Universe. One of them is a violation of $C$ (charge conjugation symmetry) and $CP$ (the composition of parity and $C$) symmetries and this condition is important for our analysis. Namely, within methods of the quantum field theory the so-called CPT-theorem has been proved \cite{Pauli:1955ep,Lueders:1954zz,Lueders:1992dq,Streater:1989vi,Jost:1957zz,Jost:1965gt}. This theorem follows from the basic assumptions of quantum theory: (i) Relativistic invariance, (ii) The spectral condition: the eigenvalues of the energy momentum operator ${\cal P}_{\mu}$ lie in or on the plus (forward) light cone $V_{+}$, (iii) Local commutativity (called also ``microscopic causality'') \cite{Streater:1989vi,Jost:1957zz,Jost:1965gt}. This theorem states that if the conditions (i), (ii) and (iii) holds in the considered system then the combined symmetry $CPT$ must be conserved. (Here $T$ denotes time-reversal symmetry). So far no one has proved that the CPT-theorem is wrong. From this theorem it follows that the mentioned Sakharov condition can be satisfied only if the time reversal symmetry $T$ is violated. This is because then the violation of $CP$ is compensated by the violation of $T$ and the combined symmetry $CPT$ can be conserved. Thus simply when considering quantum aspects of cosmological scenarios according to the contemporary physics the observed baryon-antibaryon asymmetry in our Universe can be considered as the proof that time reversal symmetry is violated. This means that irreversible quantum processes such as the decay processes can not be considered to be contrary to the basic assumptions of cosmological models.

When analyzing temporal behavior of the decaying dark energy as a function of time $t$ the following strategies can be used. One can analyze the decay process as the quantum decay process governed by the quantum mechanical decay law ${\cal P}(t)$, or one can consider the dark energy as the energy density in the metastable false vacuum state and to analyze temporal properties of this energy density similarly to the properties of the instantaneous energy of the system in the unstable state $|\psi\rangle$.

In the case of the first strategy defining
\begin{equation}
\tilde{\rho}_{\text{de}}(t) \stackrel{\rm def}{=} \rho_{\text{de}}(t) - \rho_{\text{bare}},
\end{equation}
one has $\lim_{t \to \infty} (\rho_{\text{de}}(t) - \rho_{\text{bare}}) = 0$ and one can describe such a process assuming that (see \cite{Szydlowski:2017wlv}),
\begin{equation} \label{rho-tilde(t)}
\tilde{\rho}_{\text{de}}(t) = \tilde{\rho}_{\text{de}}(t_{0})\,{\cal P}(t) \equiv \tilde{\rho}_{\text{de}}(t_{0})\Big(|{\cal A}_{c}(t)|^{2} + \,2\,\Re\,[{\cal A}_{c}(t)\,( {\cal A}_{L}(t) )^{\ast}]\,+\,|{\cal A}_{L}(t)|^{2}\Big),
\end{equation}
where ${\cal P}(t)$ is given by the relation (\ref{Ac+AL-2}). Such an approach can be considered as a generalization of the idea studied by Shafieloo {\em et al.} in \cite{Shafieloo:2016bpk}. (This was done in \cite{Szydlowski:2017wlv}).

This approach described shortly above is self-consistent if we identify $\rho_{\text{de}}(t_{0})$ with the energy $E_{0}$ of the unstable system divided by the volume $V_{0}$ (where $V_{0}$ is the volume of the system at $t=t_{0}$): $\rho_{\text{de}}(t_{0}) \equiv \rho_{\text{de}}^{\text{qft}} \stackrel{\rm def}{=} \rho_{\text{de}}^{0} = \frac{E_{0}}{V_{0}}$ and $ \rho_{\text{bare}} =\frac{E_{\text{min}}}{V_{0}}$ and $\gamma_{0} = \frac{\Gamma_{0}}{V_{0}}$. Here $\rho_{\text{de}}^{\text{qft}}$ is the vacuum energy density calculated using quantum field theory methods. In such a case
\begin{equation}
\beta = \frac{E_{0} - E_{\text{min}}}{{\Gamma}_{0}} \equiv \frac{\rho_{\text{de}}^{0} - \rho_{\text{bare}}}{{\gamma}_{0}} > 0, \label{beta-rho}
\end{equation}
(where $\gamma_{0} = {\Gamma}_{0}/V_{0}$), or equivalently, $ {\Gamma}_{0}/V_{0} \equiv \frac{\rho_{\text{de}}^{0} - \rho_{\text{bare}}}{\beta}$.

Regarding the second strategy, as it was said earlier, the general properties of the quantum decay process shortly described above do not depend on the mechanism responsible for a decay. The inherent mechanism of quantum decay processes is the reduction of the energy of the system in the metastable state $|\psi\rangle$ from the large values measured at exponential decays time region to much smaller values in the asymptotically late times region. Our hypothesis is that this pure quantum effect may take place also in the case of decaying dark energy.

One can meet in the large literature cosmological models with metastable vacuum (see, eg. \cite{Rubio:2015zia,Kennedy:1980cj,Branchina:2013jra,Branchina:2014rva} and many others). Some of these models admit the lifetime of the Universe very small \cite{Kennedy:1980cj} or even smaller than the Planck time (see \cite{Branchina:2013jra,Branchina:2014rva}). In such a case the formalism described in this section is fully applicable. Let us consider now a cosmological scenario in which false vacuum may decay at the inflationary stage of the Universe. It corresponds to the hypothesis formulated by Krauss and Dent \cite{Krauss:2007rx,Krauss:2008pt} which suggests that some false vacuum regions do survive well up to the time $T_\text{q-c}$ or later. So, let $|\psi\rangle = | 0\rangle^{\text{F}}$ be a false and $|0\rangle^{\text{T}}$ true vacuum states, respectively, and $E_{0} = E^{\,\text{false}}_{0}$ be the energy of a state corresponding to the false vacuum measured at the canonical decay time, which leads to the vacuum energy density calculated using quantum field theory methods. Let $E^{\,\text{T}}_{0}$ be the energy of true vacuum (i.e., the true ground state of the system).

As it is seen from the results presented earlier, the problem is that the energy of those false vacuum regions which survived up to $T_\text{q-c}$ and much later differs from $E^{\,\text{F}}_{0}$. Now, if one assumes that $E^{\,\text{T}}_{0} \equiv E_{\text{min}}$ and $E_{0}^{\,\text{F}} = E_{0}$ and takes into account results described above then one can conclude that the energy of the system in the false vacuum state has the following form at asymptotically late times for $t \gg T_\text{q-c}$.
\begin{equation}
{E}^{\,\text{F}}(t) \simeq E^{\,\text{T}}_{0} +
\frac{f_{2}\,\hbar^{2}}{t^{2}} + \frac{f_{4}\,\hbar^{4}}{t^{4}} + \cdots \, \neq\,E^{\,\text{F}}_{0},
\label{E-false-infty}
\end{equation}
Now, if we identify $\rho_{\text{de}}(t_{0})$ with the energy $E_{0}$ of the unstable system divided by the volume $V_{0}$ (where $V_{0}$ is the volume of the system at $t=t_{0}$): $\rho_{\text{de}}(t_{0}) \equiv \rho_{\text{de}}^{\text{qft}} \stackrel{\rm def}{=} \rho_{\text{de}}^{0} = \frac{E_{0}}{V_{0}}$ and $ \rho_{\text{bare}} =\frac{E_{\text{min}}}{V_{0}}$, (where $\rho_{\text{de}}^{\text{qft}}$ is the vacuum energy density calculated using quantum field theory methods), then at times $t \gg T_\text{q-c}$,
\begin{equation}
\rho_{\text{de}}(t) = \rho_{0}^{\text{F}}(t) \simeq \rho_{\text{bare}} + \frac{d_{2}}{t^{2}} + \frac{d_{4}}{t^{4}} + \cdots, \quad (t \gg T_\text{q-c}), \label{rho}
\end{equation}
where $d_{2k} = d_{2k}^{\ast}$.
The analogous relation takes place for $\Lambda (t) = \frac{8\pi G}{c^{2}}\,\rho(t)$, or $\Lambda (t) = 8\pi G\,\rho(t)$ in $\hbar = c =1$ units
\begin{equation}
\Lambda(t) \simeq \Lambda_{\text{bare}} + \frac{\alpha_{2}}{t^{2}} + \frac{\alpha_{4}}{t^{4}} + \cdots,\quad (t \gg T_\text{q-c}).\label{lambda3}
\end{equation}
The good approximation of eq.~(\ref{lambda3}) is to replace the cosmological time $t$ in it with the Hubble cosmological scale time $t_\text{H}=\frac{1}{H}$. As the result, instead of (\ref{lambda3}) one gets
\begin{equation}
\Lambda(t)= \Lambda(H(t)) \simeq \Lambda_{\text{bare}} + \alpha_{2} \left(H(t)\right)^{2} +
\alpha_{4} \left(H(t)\right)^{4} + \cdots \, , \label{L(H)}
\end{equation}
that is exactly the parameterization considered in \cite{Shapiro:2003ui,EspanaBonet:2003vk} and in many papers of these and other authors.

From the above analysis it follows that $\Lambda_{\text{bare}}$ is the limiting value of $\Lambda(t)$ reached when $t \to \infty$. For $T_{q-c}$ smaller or much smaller than the age of the Universe ${\cal T}_{0}$, i.e. for $T_{\text{q-c}} \ll {\cal T}_{0}$ the reasonable approximation is to assume that $\Lambda_{\text{bare}}$ equals to the present measured value of the cosmological constant and to use the relation $\Lambda_{0} \simeq 10^{120} \Lambda_{\text{bare}}$, where $\Lambda_{0} = \Lambda_\text{qft} = \frac{8\pi G}{c^{2}}\,\rho_{\text{de}}^{0}$. As it was mentioned earlier $ \rho_{\text{bare}} =\frac{E_{\text{min}}}{V_{0}}$ and this means if we knew the form of the Hamiltonian $ \mathfrak{H}$ describing the quantum processes acting in the early Universe then we would know its spectrum and so $E_{\text{min}}$ and thus assuming $V_{0}$ we would knew the density $ \rho_{\text{bare}}$ and thus the value of $\Lambda_{\text{bare}} \neq 0$. However, without knowing the exact form of $ \mathfrak{H}$ and knowing that such a Hamiltonian must exist, within the use of our method we know at the same time that there must be some $\Lambda_{\text{bare}} \neq 0$ and generally the approximate value of $\Lambda_{\text{bare}} $ and other parameters can be fitted form the observational data. Summing up, the emergence of the $\Lambda_{\text{bare}}$ is the result of the quantum process of the creation of the Universe.

If we, for example, postulate a Hamiltonian (the energy operator $\mathfrak{H}$) for a homogeneous self-interacting scalar field with the potential $V(\phi)$ then using the Fock-Krylov approach we can obtain the quantum description of tunneling effect similarly as it was done, e.g. in the paper by Calzetta and Verdaguer \cite{Calzetta:2006rg} and other papers (e.g. \cite{Winter:1961zz,vanDijk:1999zz,vanDijk:2002ru,Dicus:2002tdq,Martorell:2009qpd,vanDijk:2019}).

Note that the form of $\kappa (t)$ does not change when one passes from energies $E(t)$, $E_{0}$, $E_\text{min}$ to the above defined energy density $\rho$ and $\Lambda$
\begin{align}
\kappa (t) = \frac{E(t) - E_{\text{min}}}{E_{0} - E_{\text{min}}}
\equiv \frac{\frac{E(t)}{V_{0}} - \frac{E_{\text{min}}}{V_{0}}}{\frac{E_{0}}{V_{0}} - \frac{E_{\text{min}}}{V_{0}}}
= \frac{\rho_{\text{de}}(t) - \rho_{\text{bare}}}{\rho_{\text{de}}^{0} - \rho_{\text{bare}}} = \frac{\Lambda (t) - \Lambda_{\text{bare}}}{\Lambda_{0} - \Lambda_{\text{bare}}}.
\label{kappa-L}
\end{align}

\section{Cosmological implications of decaying vacuum}

Our model has a methodological status of an effective theory which has a cut-off on the scale time because it describes the evolution of the universe only for times when dark energy is parameterized by the power series with respect to $1/t$, not for earlier times. For $t>T_{q-c}$, the decay of the metastable vacuum is under the power-law regime. Now it will be convenient to estimate the order of this parameter from the best fit values of model parameters. From the statistical analysis obtained in ref.~\cite{Szydlowski:2015fya}, we know the value of the best fit of the parameters $\Omega_{\alpha^2,0}=\frac{\alpha_2}{3H_0^2}$, where $\Omega_{\alpha^2,0}=-0.000210$ and $H_0=68.38\frac{\text{km}}{\text{s Mpc}}$. In result, we obtain that the best fit of the parameter $\beta$ is equal to 0.000234. Next, substituting the best fit value of the parameter $\beta$ and $E_0=10^{120} E_\text{min}$ into eq.~(\ref{Tqc1}) we obtain that the time $T_\text{q-c}$ equals $T_\text{q-c} \approx 10^{-44}\,\text{s}$. This value corresponds to a period after the Planck epoch.

Parameterization (\ref{E(t)-1}) of the energy density of dark energy is valid when the adiabatic condition $\frac{|\dot E(t)|}{|E(t)|}\gg\frac{\dot a}{a}$ (or $\frac{|\dot \rho_\text{de}|}{|\rho_\text{de}|}\gg\frac{\dot a}{a}$) is achieved, where $a$ is the scale factor. The physical sense of this requirement is that scale of quantum processes of decaying vacuum is much smaller than the cosmological time scale. The satisfaction of this condition allows us to apply the perturbation methods to calculate quantum effect on the background of an almost static spacetime. From this condition we get that the adiabatic condition is violated when $\beta>10^{7}$ or $E_0>10^{93}E_\text{min}$ ( $\rho_\text{de}^0>10^{93}\Lambda_\text{bare}$), where $\rho_\text{de}^0$ is the initial value of the energy density of dark energy. Therefore if $\rho_\text{de}^0<10^{93}\Lambda_\text{bare}$ and $\beta<10^{7}$ then the adiabatic condition can be achieved during the quantum process of the decay of the vacuum. This conditions are satisfied in the very early universe. Therefore, our model is an effective description of decay of quantum vacuum in this restricted era.

The parameter $\beta$ characterizes the type of particles. It is very large for neutrons ($\beta = 1.72951 \cdot 10^{24}$) while it is in some interval with the upper limit of $\beta < 31325$ ($E_\text{min} > 0$) \cite{Tanabashi:2018oca}.

\begin{figure}
\centering
\includegraphics[width=0.7\linewidth]{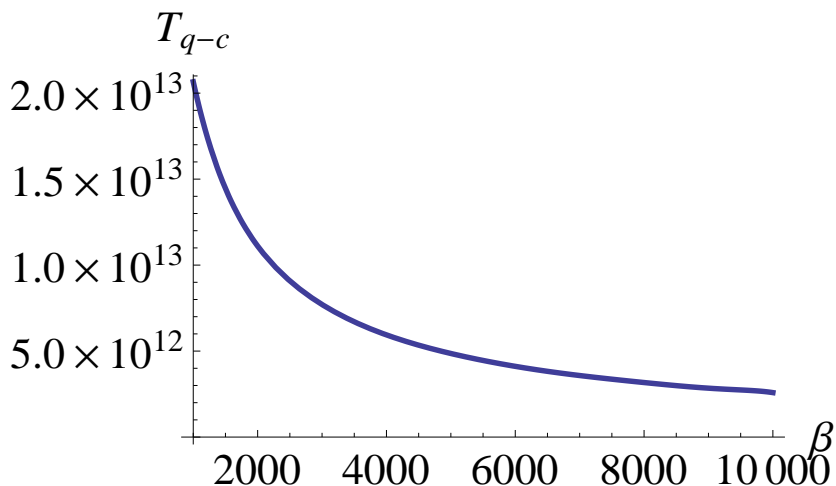}
\caption{The dependence between the time $T_\text{q-c}$ and the value of $\beta$ parameter for $\rho_\text{de}^0=10^{85}\Lambda_\text{bare}$. The unit of $T_\text{q-c}$ is expressed in seconds.}
\label{fig11}
\end{figure}

The adiabatic epoch takes place in our model during the early stage of the evolution of the Universe provided that the upper bound for $\beta$ is $\beta < 10^7$. The parameter $\beta$ characterizes hypothetical particles, from which the quantum vacuum is build. The quantum vacuum in our model is a metastable state, which decays with increasing time $t$ and quantum behavior of this vacuum becomes practically negligible for times $t > T_\text{q-c}$ as it was shown in section~3 (see figures~\ref{f1-all}, \ref{f2-all} and \ref{f-E}). As a result the classical vacuum emerges. It should be stressed that as one can see in figure~\ref{f-E} the density of the dark energy $\rho_\text{de}(t)$ at times $t \ll T_\text{q-c}$ is much, much larger then at times $t \gg T_\text{q-c}$, oscillates for $t \lesssim T_\text{q-c}$ and rapidly decreases to very, very small values for $t \gg T_\text{q-c}$ (see formula (\ref{kappa-L}) and figure~\ref{f-E}). The time $T_\text{q-c}$ depends on the values of the parameter $\beta$, and the density of dark energy $\rho^{0}_\text{de}$. In figure~\ref{fig11} one can see changes of the time $T_\text{q-c}$ depending on the values of the parameter $\beta$, $T_\text{q-c} = T_\text{q-c}(\beta)$, obtained numerically. Calculations were performed for $\rho_\text{de}^0=10^{85} \Lambda_\text{bare}$. Note that $T_\text{q-c}(\beta)$ decreases as $\beta$ grows.

We consider cosmological equations with $\rho_\text{de}=\Lambda$, which describes the density of dark energy. Here, the form of $\Lambda$ has been derived in the previous section and is given by
\begin{equation}
\Lambda \equiv \Lambda_{\text{eff}}(t) = \Lambda_\text{bare}+\delta\Lambda(t),\label{lambda2}
\end{equation}
where $\delta \Lambda(t)$ describes quantum corrections given by a series with respect to $\frac{1}{t}$, i.e.,
\begin{equation}
\delta\Lambda(t)=\sum^\infty_{n=1}\alpha_{2n}\left(\frac{1}{t}\right)^{2n},\label{sum}
\end{equation}
where $t$ is the cosmological scale time. The function $\delta \Lambda(t)$ has a reflection symmetry with respect to the cosmological time ($\delta \Lambda(-t)=\delta \Lambda(t)$).

The Friedmann equation is given by the following equation
\begin{equation}
3H(t)^2=\rho_\text{m}(t)+\rho_\text{de}(t) \label{friedmann}
\end{equation}
where $H(t)$ is the Hubble function.

We assume the equation of state for matter and dark energy in the form $p_\text{m}=w_\text{m}\rho_\text{m}$ and $p_\text{de}=-\rho_\text{de}$, respectively, where $p_\text{m}$ is pressure of matter and $p_\text{de}$ is pressure of dark energy. The Einstein field equation for the FRW metric is given by
\begin{equation} \label{hubble}
\begin{split}
\frac{dH(t)}{dt} &=-\frac{1}{2}(\rho_\text{eff}(t)+p_\text{eff}(t))=-\frac{1}{2} (w_\text{m}+1)\rho_\text{m}(t) \\ &= -\frac{1}{2}(w_\text{m}+1)\left(3H(t)^2-\Lambda_\text{bare}-\delta \Lambda(t)\right),
\end{split}
\end{equation}
where $\rho_\text{m}$ is the density of matter, $\rho_\text{eff}=\rho_\text{m}+\rho_\text{de}$ and $p_\text{eff}=p_\text{m}+p_\text{de}$.

The total energy-momentum tensor $T^{\mu\nu}=T^{\mu\nu}_\text{m}+T^{\mu\nu}_\text{de}$ is conserved in our model. In this model, the energy density is transferred between the matter and dark energy, what gives the interaction in the dark sector. This process is described by the following equations
\begin{equation}
\begin{split}
\frac{d\rho_\text{m}(t)}{dt}+3H(t)\rho_\text{m}(t) & =-\frac{d\rho_\text{de}(t)}{dt}=-Q(t),\\
\frac{d\rho_\text{de}(t)}{dt} & =Q(t),\label{rho2}
\end{split}
\end{equation}
where it is assumed that pressure of matter $p_\text{m}=0$ and $p_\text{de}=-\rho_\text{de}$. We obtain eqs~(\ref{rho2}) from eq.~(\ref{friedmann}) and (\ref{hubble}). In eq.~(\ref{rho2}), an interacting term appears, which is defined as $Q(t)=\frac{d\rho_\text{de}(t)}{dt}=\frac{d\Lambda_{\text{eff}}(t)}{dt}$. The interacting term is non-zero in the case when energy flows between dark energy and dark matter.

\begin{figure}
\centering
\includegraphics[width=0.7\linewidth]{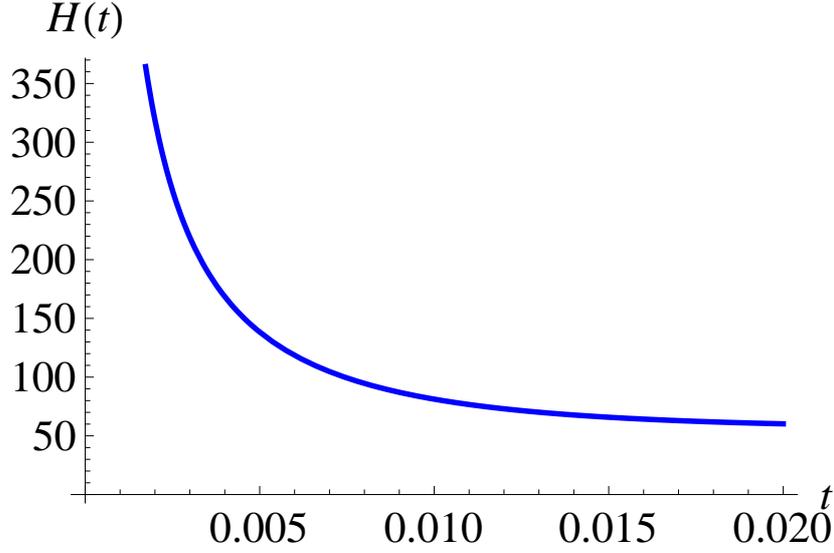}
\caption{The diagram presents the typical evolution of the Hubble function $H(t)$ for the model with $\rho_\text{de}=\Lambda_\text{bare}+\frac{\alpha_{4}}{t^4}$ and dust matter for the case $\alpha_4=10^{-8}\frac{\text{s}\ \text{Mpc}}{\text{km}}$. The cosmological time $t$ is expressed in $\frac{\text{s}\ \text{Mpc}}{\text{km}}$ and $H(t)$ is expressed in $\frac{\text{km}}{\text{s Mpc}}$.}
\label{fig10}
\end{figure}

Now we cut off the series (\ref{sum}) on the second term. In result, we get
\begin{equation}
\delta\Lambda(t)=\frac{\alpha_{2}}{t^{2}}+\frac{\alpha_{4}}{t^{4}}.\label{series}
\end{equation}
From eq.~(\ref{Re-h-as}), we can obtain that the parameter $\alpha_2$ is negative and the parameter $\alpha_4$ is negative when $\beta\in(0.346, 0.684)$.
In this case, the Friedmann equation has the following form
\begin{equation}
3H^2=\rho_\text{m}+\Lambda_\text{bare}+\frac{\alpha_{2}}{t^{2}}+\frac{\alpha_{4}}{t^{4}}.\label{friedmann3}
\end{equation}
The typical evolution of the Hubble function is presented in figure~\ref{fig10}.

The Friedmann equation (\ref{friedmann3}) can be rewritten in dimensionless terms
\begin{equation}
\frac{H(t)^2}{H_0^2}=\Omega_\text{m,0}f(t)+\Omega_{\Lambda_\text{bare},0}+\Omega_{\alpha_{2},0}\frac{{\cal T}_{0}^2}{t^2}+\Omega_{\alpha_{4},0}\frac{{\cal T}_{0}^4}{t^4}, \label{eq:FEdimless}
\end{equation}
where $\Omega_\text{m,0}=\frac{\rho_\text{m,0}}{3H_0^2}$, $\Omega_{\Lambda_\text{bare},0}=\frac{\Lambda_\text{bare}}{3H_0^2}$, $\Omega_{\alpha_{2},0}=\frac{\alpha_\text{2}}{3H_0^2 {\cal T}_{0}^2}$, $\Omega_{\alpha_{4},0}=\frac{\alpha_\text{4}}{3H_0^2 {\cal T}_{0}^4}$, $H_0$ is the present value of the Hubble constant, $f(t)=\frac{\rho_\text{m}(t)}{\rho_\text{m,0}}$, and ${\cal T}_{0}$ is the present age of the Universe. Here, $\rho_\text{m,0}$ parameter denotes the present value of the energy density of matter.

Our model has five parameters: $H_0$, $\Omega_\text{m,0}$, $\Omega_{\Lambda_\text{bare},0}$, $\Omega_{\alpha_{2},0}$, and $\Omega_{\alpha_{4},0}$. But, for the present age of the Universe, from eq.~(\ref{eq:FEdimless}), we obtain the following condition
\begin{equation}
1=\Omega_\text{m,0}+\Omega_{\Lambda_\text{bare},0}+\Omega_{\alpha_{2},0}+\Omega_{\alpha_{4},0},
\end{equation}
which reduces the number of independent parameters from five to four.

In the case $\delta\Lambda(t)$ given by eq.~(\ref{friedmann3}), when $\Lambda_\text{bare}=0$, eq.~(\ref{hubble}) has the following solution
for $\alpha_{4}>0$
\begin{equation} \label{bessel1}
H(t)=\frac{1-\sqrt{1+3(w_\text{m}+1)^2 \alpha_2}}{3(w_\text{m}+1)t} -\frac{\sqrt{\alpha_4}I_{n+1}\left(\frac{(w_\text{m}+1)\sqrt{3\alpha_4}}{2t}\right)}{\sqrt{3}t^2 I_{n}\left(\frac{(w_\text{m}+1)\sqrt{3\alpha_4}}{2t}\right)},
\end{equation}
for $\alpha_{4}<0$
\begin{equation} \label{bessel2}
H(t)=\frac{1-\sqrt{1+3(w_\text{m}+1)^2 \alpha_2}}{3(w_\text{m}+1)t} + \frac{\sqrt{-\alpha_4}J_{n+1}\left(\frac{(w_\text{m}+1)\sqrt{-3\alpha_4}}{2t}\right)}{\sqrt{3}t^2 J_{n}\left(\frac{(w_\text{m}+1)\sqrt{-3\alpha_4}}{2t}\right)},
\end{equation}
where $I_{n}(x)$ is the modified Bessel function of the first kind, $J_{n}(x)$ is the Bessel function of the first kind and $n=\frac{1}{2}\sqrt{1+3(w_\text{m}+1)^2\alpha_2}$.

Note that for $\alpha_4=0$, eqs.~(\ref{bessel1}) and (\ref{bessel2}) simplify to the following form
\begin{equation}
H(t)=\frac{1-\sqrt{1+3(w_\text{m}+1)^2\alpha_\text{2}}}{3t}. \label{eq:57}
\end{equation}
From eq.~(\ref{eq:57}), we obtain a formula for the scale factor $a(t)$
\begin{equation}
a(t)=\left(\frac{t}{{\cal T}_{0}}\right)^{\frac{1}{3}\left(1-\sqrt{1+3(w_\text{m}+1)^2\alpha_{2}}\right)},
\end{equation}
where ${\cal T}_{0}$ is the present age of the Universe.

For the illustration of dynamical analysis of cosmology with dark energy parameterization (\ref{sum}), we choose a special case in which $\alpha_{2}<0$. Let $x=\frac{\sqrt{\rho_\text{m}}}{\sqrt 3H}$, $y=\frac{\sqrt{\Lambda_\text{bare}}}{\sqrt 3H}$ and $z=\frac{\sqrt{-\alpha_{2}}}{\sqrt 3Ht}$. From the Friedmann equation (\ref{friedmann3}), we get the following condition
\begin{equation}
1=x^2+y^2-z^2+\sum^\infty_{n=2}\frac{\alpha_{(2n)}\Lambda^{n-1}z^{2n}}{\sqrt{-\alpha_2}^{2n}y^{2n-2}}.\label{condition}
\end{equation}
We choose a new reparameterized time $\tau=\ln a(t)$, where $a(t)$ is the scale factor and $\frac{\dot{a}(t)}{a(t)} = H(t)$. We obtain $y'$ and $z'$, where $'\equiv\frac{d}{d\tau}$, and get the dynamical system
\begin{equation}
\begin{split}
y' &=\frac{3}{2}y x^2,\\
z' &=\frac{3}{2}z x^2-\sqrt\frac{-3}{\alpha_{2}}z^2. \label{eq:ds60}
\end{split}
\end{equation}
We can use condition (\ref{condition}) to eliminate the variable $x$ from the dynamical system (\ref{eq:ds60}). In result, we get the following dynamical system
\begin{equation}
\begin{split}
y' &=\frac{3}{2}y \left(1-y^2-\sum^\infty_{n=1}\frac{\alpha_{2n}\Lambda^{n-1}z^{2n}}{\sqrt{-\alpha_2}^{2n}y^{2n-2}}\right),\\
z' &=\frac{3}{2}z \left(1-y^2-\sum^\infty_{n=1}\frac{\alpha_{2n}\Lambda^{n-1}z^{2n}}{\sqrt{-\alpha_2}^{2n}y^{2n-2}}\right)-\sqrt\frac{-3}{\alpha_{2}}z^2.
\end{split}
\end{equation}

\begin{figure}
\centering
\includegraphics[width=0.7\linewidth]{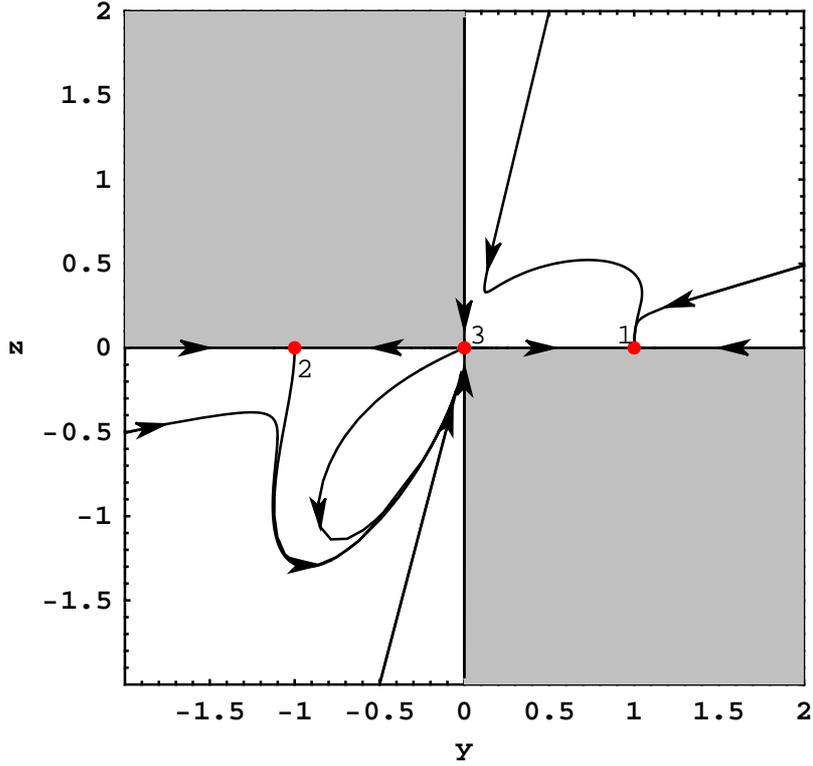}
\caption{The phase portrait of the system (\ref{dynamical3}). Critical point 1 $\left(y=1,\ z=0\right)$ is a stable node and represents the de Sitter universe. Critical point 2 $\left(y=-1,\ z=0\right)$ is a saddle and represents the anti de Sitter universe. Critical point 3 $\left(y=0,\ z=0\right)$ is a saddle and represents the Einstein-de Sitter universe. The values of $\alpha_{2}$ and $\gamma$ are assumed respectively as $-1$ and $1$. The gray color represents the non-physical domain forbidden for classical trajectories for which $H^2\geq 0$.}
\label{fig6}
\end{figure}

If we cut off the series (\ref{sum}) on the second term ($n=2$) then we get
\begin{equation}
\begin{split}
\frac{dy}{d\tau} &=\frac{3}{2}y \left(y^2(1-y^2+z^2)-\gamma z^4\right),\\
\frac{dz}{d\tau} &=\frac{3}{2}z \left(y^2(1-y^2+z^2)-\gamma z^4\right)-\sqrt\frac{-3}{\alpha_{2}}y^2 z^2,\label{dynamical3}
\end{split}
\end{equation}
where $\gamma=\frac{\alpha_4 \Lambda}{\alpha^2_2}$ and $d\tau=\frac{dt}{y^2}$ is the new reparameterized time.
The phase portrait for system (\ref{dynamical3}) is presented in figure~\ref{fig6}.

For the analysis of the behavior of trajectories at infinity we use the following sets of projective coordinates
\begin{enumerate}
\item $Y=\frac{1}{y}$, $Z=\frac{z}{y}$,
\item $\tilde Y=\frac{y}{z}$, $\tilde Z=\frac{1}{z}$.
\end{enumerate}

\begin{figure}
\centering
\includegraphics[width=0.7\linewidth]{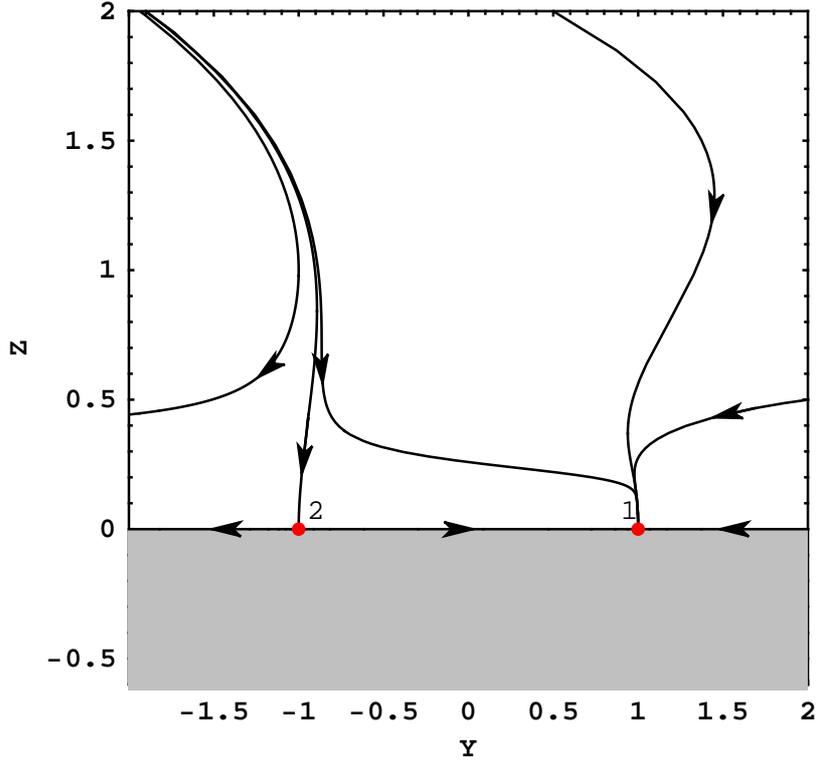}
\caption{The phase portrait of the system (\ref{dynamical3}). Critical point 1 $\left(Y=1,\ Z=0\right)$ is a stable node and represents the de Sitter universe. Critical point 2 $\left(Y=-1,\ Z=0\right)$ is a saddle and represents the anti de Sitter universe. The values of $\alpha_{2}$ and $\gamma$ are assumed respectively as $-1$ and $1$. The gray color represents the non-physical domain forbidden for classical trajectories for which $H^2\geq 0$.}
\label{fig1}
\end{figure}

For the coordinates $Y$ and $Z$ with the reparameterized time $d\tau=\frac{dt}{Y}$, we get the following dynamical system
\begin{equation}
\begin{split}
\frac{dY}{d\tau} &=\frac{3}{2}\left(1-Y^2-Z^2+\gamma Z^4\right),\\
\frac{dZ}{d\tau} &=-\frac{\sqrt{3}Z^2}{\sqrt{-\alpha_2}}.\label{dynamical4}
\end{split}
\end{equation}
The phase portrait for system (\ref{dynamical4}) is presented in figure~\ref{fig1}.

\begin{figure}
\centering
\includegraphics[width=0.7\linewidth]{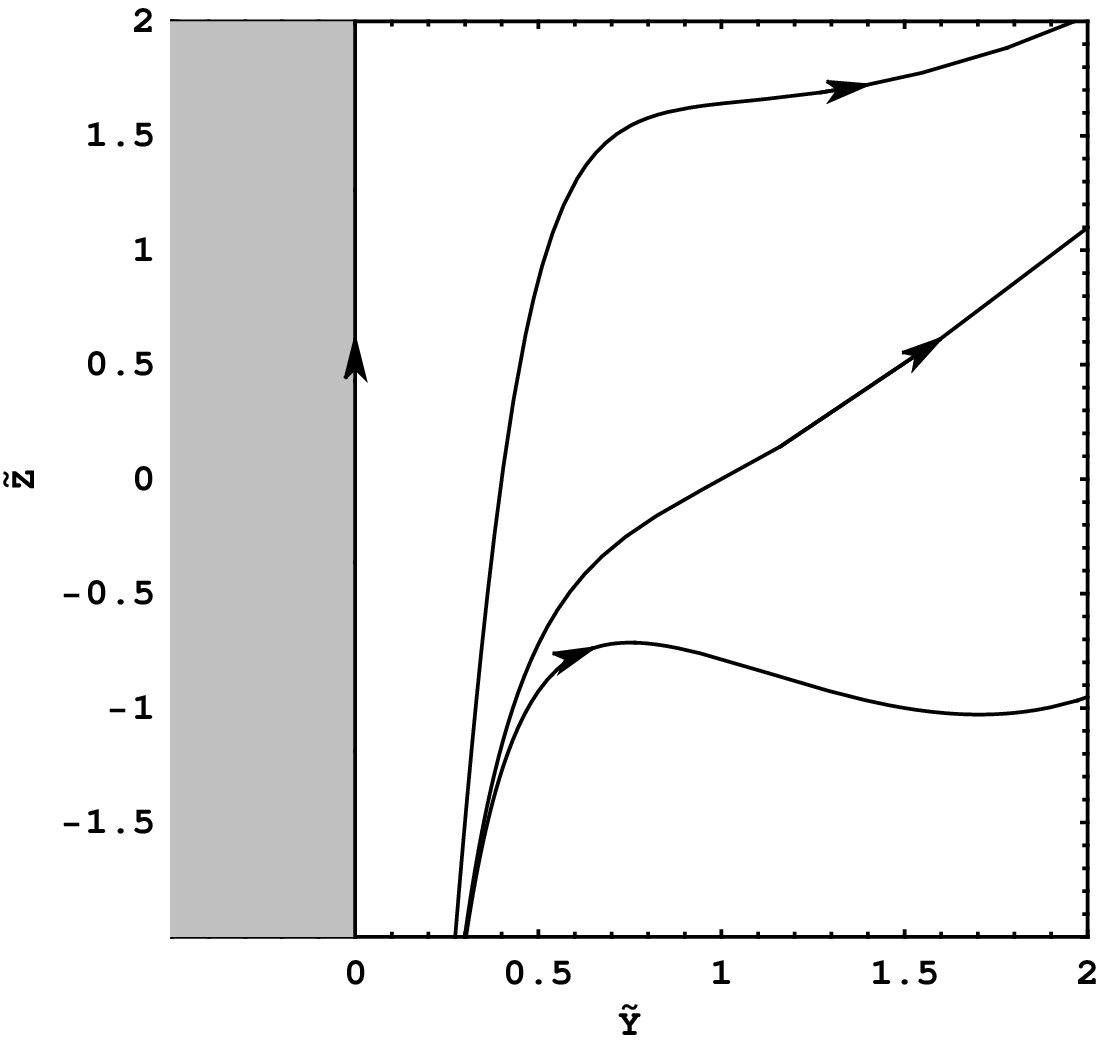}
\caption{The phase portrait of the system (\ref{dynamical5}). The values of $\alpha_{2}$ and $\gamma$ are assumed as $-1$ and $1$, respectively. The gray color represents the non-physical domain forbidden for classical trajectories for which $H^2\geq 0$.}
\label{fig9}
\end{figure}

For coordinates $\tilde Y$, $\tilde Z$, we obtain the dynamical system
\begin{equation}
\begin{split}
\frac{d\tilde Y}{d\tau} &=\frac{\sqrt{3}\tilde Y^3}{\sqrt{-\alpha_2}},\\
\frac{d\tilde Z}{d\tau} &=\frac{\sqrt{3}}{\sqrt{-\alpha_2}}\tilde Y^2\tilde Z+3\left(\gamma+\tilde Y^4-\tilde Y^2 (1+\tilde Z^2)\right),\label{dynamical5}
\end{split}
\end{equation}
where the reparameterized time $d\tau=\frac{dt}{\tilde Y^2 \tilde Z}$. The phase portrait for this system is presented in figure~\ref{fig9}.

In parameterization (\ref{sum}) the symmetry $t\rightarrow -t$ is present. This symmetry appears in the acceleration equation
\begin{equation}
\frac{dH(t)}{dt} =-\frac{1}{2}\left(3H(t)^2-\Lambda_\text{bare}-\sum^\infty_{n=1}\alpha_{2n}t^{-2n}\right).\label{hubble2}
\end{equation}

\begin{figure}
\centering
\includegraphics[width=0.7\linewidth]{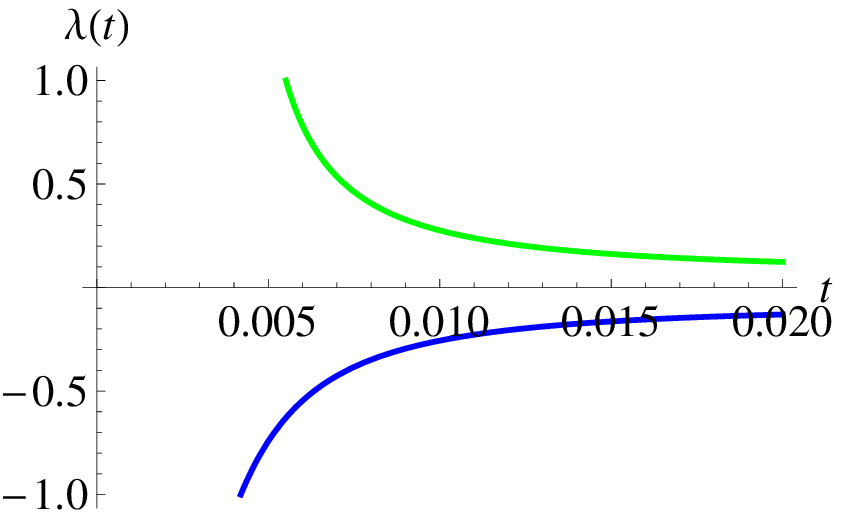}
\caption{The diagram presents the evolution of the parameter $\lambda(t)$ with respect to the cosmological time $t$ for $\rho_\text{de}=\Lambda_\text{bare}+\frac{\alpha_{4}}{t^4}$ for the dust matter case. Two example values of the parameter $\alpha_4$ are chosen: $\alpha_4=-10^{-7}\frac{\text{s}\ \text{Mpc}}{\text{km}}$ (the top green curve) and $\alpha_4=10^{-7}\frac{\text{s}\ \text{Mpc}}{\text{km}}$ (the bottom blue curve). The cosmological time $t$ is expressed in $\frac{\text{s}\ \text{Mpc}}{\text{km}}$. Here, the present age of the Universe ${\cal T}_0=0.014\frac{\text{s}\ \text{Mpc}}{\text{km}}$.}
\label{fig4}
\end{figure}

The small deviation from the canonical scaling of the energy density of matter $\rho_\text{m}\propto a^{-3(1+w_\text{m})}$ \cite{Shapiro:2003ui,EspanaBonet:2003vk} is defined by the following term
\begin{equation}
\lambda(t)=\frac{\ln\frac{\rho_\text{m}(t)}{\rho_\text{m,0}}}{\ln a(t)}+3(1+w_\text{m})\label{deviation}
\end{equation}
in relation
\begin{equation}
\rho_\text{m}(t)=\rho_\text{m,0}a^{-3(1+w_\text{m})+\lambda(t)}.
\end{equation}
The evolution of the $\lambda(t)$ parameter is presented in figure~\ref{fig4}. For radiation, we have an analogical relation
\begin{equation}
\rho_\text{r}=\rho_\text{r,0}a^{-4+\lambda(t)}.
\end{equation}
Equation~(\ref{deviation}) can be rewritten in the form
\begin{equation}
\lambda(t)=\frac{1}{\ln a(t)}\int \frac{\dot \Lambda(t)}{H(t)\rho_\text{m}(t)} d\ln a.
\end{equation}
If we cut off the series (\ref{sum}) on the first term then $\lambda(t)$ is limited by $|\lambda(t)|\geq 0$.

\section{Decay as a result of an irreversible quantum process determining the arrow of time}

In the cosmological studies, one can associate the Hawking temperature and entropy with the apparent horizon in an analogous way to as it is considered in the context of the black hole horizon \cite{Cai:2005ra, Gibbons:1977mu}. While in the de Sitter space-time, the event horizon coincides with the apparent horizon of the FRW flat cosmological model, for more general cosmological models, the apparent horizon related with the Hawking temperature and entropy always exists even if the horizon does not exist.

In this section, we study thermodynamics in the model under consideration which can be reduced to the problem of thermodynamics in cosmological models with the matter (dark matter and baryonic matter) and dark energy interaction. This problem was investigated in the context of the generalized second law of thermodynamics in interacting cosmological models \cite{Jamil:2009eb, Karami:2009yd}.

In our model, energy densities of both matter and dark energy satisfy the continuity equation in the form
\begin{gather}
\frac{d}{dt}\left(\rho_\text{m}+\rho_\text{de}\right)+3H\left(\rho_\text{m}+p_\text{m}+\rho_\text{de}+p_\text{de}\right)=0, \label{eq:68}\\
\frac{d}{dt}\rho_\text{de}=\frac{d}{dt}\left(\delta \Lambda(t)\right). \label{eq:69}
\end{gather}

If we assume that matter is in the form of dust ($p_\text{m}=0$) and dark energy of the vacuum satisfies the equation of state $p_\text{vac}=-\rho_\text{vac}$, then equations (\ref{eq:68}) and (\ref{eq:69}) assume the form of the following equation for both constituents
\begin{align}
\dot\rho_\text{m}+3H\rho_\text{m}&=-\frac{d}{dt}\left(\delta\Lambda(t)\right), \label{eq:70} \\
\dot\rho_\text{de}&=\frac{d}{dt}\left(\delta \Lambda(t)\right), \label{eq:71}
\end{align}
where $\delta\Lambda(t)$ is an additive and time dependent contribution added to the bare cosmological constant (see eq.~(\ref{lambda2})). Let us define the interacting term $Q\equiv\frac{d}{dt}\left(\delta \Lambda(t)\right)$. It would be convenient to rewrite eqs.~(\ref{eq:70}) and (\ref{eq:71}) in a new form containing an effective coefficient equation of state $w_\text{eff}=\frac{p_\text{eff}}{\rho_\text{eff}}$, i.e.,
\begin{align}
\dot\rho_\text{de}+3H\left(1+w^\text{eff}_\text{de}\right)\rho_\text{de}=0,\\
\dot\rho_\text{m}+3H\left(1+w^\text{eff}_\text{m}\right)\rho_\text{m}=0,
\end{align}
where parameters of the effective equation of state are given by
\begin{align}
w^\text{eff}_\text{de} &=-1-\frac{\frac{d}{dt}\left(\delta \Lambda(t)\right)}{3H\rho_\text{de}},\\
w^\text{eff}_\text{m} &=\frac{\frac{d}{dt}\left(\delta \Lambda(t)\right)}{3H\rho_\text{m}}.
\end{align}
After adding to above equations the Friedmann first integral $3H^2=\rho_\text{m}+\rho_\text{de}$ we obtain complete equation describing the evolution of constituents (matter and dark energy) in the background of the FRW flat cosmology. Because of the time dependence of the interacting term $Q\equiv\frac{d}{dt}\left(\delta \Lambda(t)\right)$ in the effective coefficients $w^\text{eff}_\text{de}$ and $w^\text{eff}_\text{m}$, the equation of state is also time dependent.

In the model under consideration dark energy and dark matter interact due to a quantum correction included in the model. It would be useful to investigate the thermodynamic behavior of a cosmological model with matter, dark energy, and the interaction. Therefore, we consider the universe as a thermodynamic system and apply the generalized second thermodynamic law \cite{Unruh:1982ic, Sheykhi:2008qs}. In the context of the application of thermodynamics to the Universe, the ``radius'' of the Universe is connected with the temperature of the Universe \cite{Jacobson:1995ab, Padmanabhan:2003gd}. For the flat universe as a ``radius'' is used the apparent horizon, which in this case coincides with the Hubble horizon, i.e.,
\begin{equation}
\tilde r_A=\frac{1}{|H|}\quad \text{or} \quad \frac{1}{\tilde r_A^2}=\frac{1}{3}\left(\rho_\text{de}+\rho_\text{m}\right).
\end{equation}
Now one can connect the apparent horizon (casual horizon) with the gravitational entropy and considers the Universe as a thermodynamic system for which the apparent horizon surface is its boundary.

Let us calculate a sum of the total entropy enclosed by the apparent horizon and the entropy of the apparent horizon. We assume that after the equilibrium is reached all the fluids inside the apparent horizon possess a temperature $T$ which coincides with the temperature of the horizon $T_h$ \cite{Frolov:2002va, Cai:2005ra}.

We considered the Universe as an isolated system for which the Gibbs relation is fulfilled. Therefore, the first law of thermodynamics assumes the following form
\begin{equation}
TdS=dE+PdV,
\end{equation}
where the universe filled (because of the nonzero interacting term) with effective matter and effective dark energy assumes the following forms \cite{Wang:2005pk, Gong:2006sn}
\begin{align}
dS^\text{eff}_\text{de} &=\frac{1}{T}\left(p^\text{eff}_\text{de}dV+dE_\text{de}\right),\\
dS^\text{eff}_\text{m} &=\frac{1}{T}\left(p^\text{eff}_\text{m}dV+dE_\text{m}\right),
\end{align}
where volume $V=\frac{4\pi}{3}\tilde r_A^3$ represents the volume limited by the apparent horizon (from this we obtain $dV=4\pi\tilde r_A^2 d\tilde r_A$, $E_\text{m}=\rho_\text{m}V$ and $E_\text{de}=\rho_\text{de}V$. After dividing above formulas over $dt$ we obtain relation for the time variability of entropies of both effective matter and dark energy
\begin{align}
p^\text{eff}_\text{m}=w^\text{eff}_\text{m}\rho_\text{m},\\
p^\text{eff}_\text{de}=w^\text{eff}_\text{de}\rho_\text{de}.
\end{align}
Finally, the following result can be obtained \cite{Jamil:2009eb}
\begin{align}
\frac{dS^\text{eff}_\text{de}}{dt}=\frac{4\pi}{T}\tilde r_A^2 \left(\dot{\tilde r}_A-H\tilde r_A\right)\left(1+w^\text{eff}_\text{de}\right)\rho_\text{de},\\
\frac{dS^\text{eff}_\text{m}}{dt}=\frac{4\pi}{T}\tilde r_A^2 \left(\dot{\tilde r}_A-H\tilde r_A\right)\left(1+w^\text{eff}_\text{m}\right)\rho_\text{m}.
\end{align}

\begin{figure}
\centering
\includegraphics[width=0.7\linewidth]{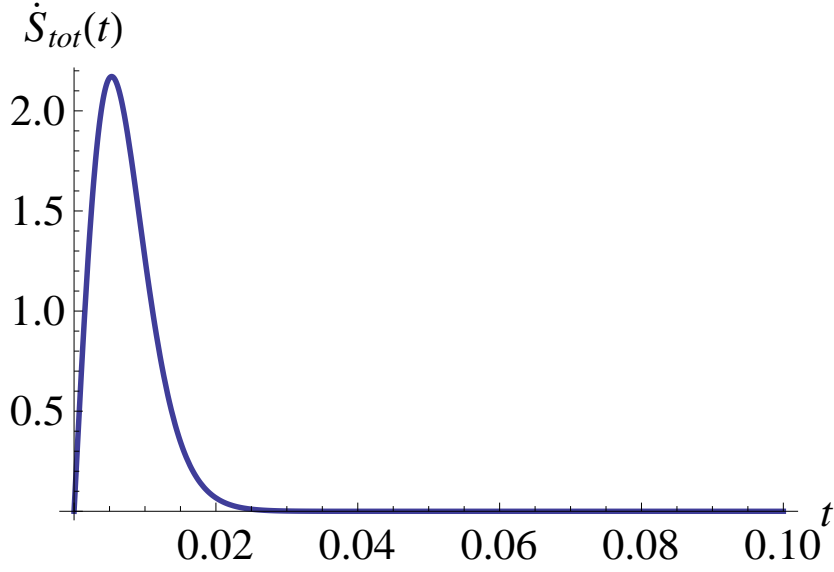}
\caption{The diagram presents the variation of the first derivative of total entropy $\dot S_\text{tot}$ with respect to the cosmological time for the case $\rho_\text{de}=\Lambda_\text{bare}+\frac{\alpha_{2}}{t^2}$, where $\alpha_\text{2}<0$. The cosmological time $t$ and $\dot S_\text{tot}$ are expressed in $\frac{\text{s}\ \text{Mpc}}{\text{km}}$. Note that $\dot S_\text{tot}$ approaches asymptotically zero as $t$ tends to the infinity.}
\label{fig2}
\end{figure}

After putting the temperature of the horizon following from black hole thermodynamics $T_\text{h}=\frac{1}{2\pi\tilde r_A}$ we obtain entropy of the horizon and its time derivatives
\begin{equation}
\frac{d S_\text{h}}{dt}=16\pi^2\tilde r_A\dot{\tilde r}_A.
\end{equation}
Because
\begin{equation}
\frac{dS_\text{tot}}{dt}=\frac{d}{dt}\left(S^\text{eff}_\text{de}+S^\text{eff}_\text{m}+S_\text{h}\right)
\end{equation}
generalized second law of thermodynamics assumes the following final result \cite{Jamil:2009eb}
\begin{equation}
\dot S_\text{tot}=4\pi^2\tilde r_A^6 H\left[(1+w_\text{de})\rho_\text{de}+(1+w_\text{m})\rho_\text{m}\right]^2,\label{entropy}
\end{equation}
where $\dot{\tilde r}_A$ is from a relation obtained after differentiating both sides of the Friedmann equation rewritten to the form
\begin{equation}
\frac{1}{\tilde r_A^2}=\frac{1}{3}\left(\rho_\text{de}+\rho_\text{m}\right).
\end{equation}
Note that the quantum effect correction is present in the interacting term $Q$, does not contribute in $\dot S_\text{tot}$ which is always non-negative. This fact confirms the validity of the generalized second law of thermodynamics. This result is obvious independent on the quantum correction arising from the decaying vacuum. The evolution of $\dot S_\text{tot}$ for the case $\rho_\text{de}=\Lambda_\text{bare}+\frac{\alpha_{2}}{t^2}$ is presented in figure~\ref{fig2}. The total entropy must always increase because $\dot S_\text{tot}\geq 0$. Effect of quantum decay of vacuum in calculation of entropy is manifested by the presence of an interacting term describing energy transfer between sectors of dark matter and dark energy. Our calculations shows that, the time derivative of the total entropy is always non-negative, and this result hold independently of the specific interaction term, i.e., the process of quantum decay is neutral with respect to the time derivative of the total entropy.

\begin{figure}
\centering
\includegraphics[width=0.7\linewidth]{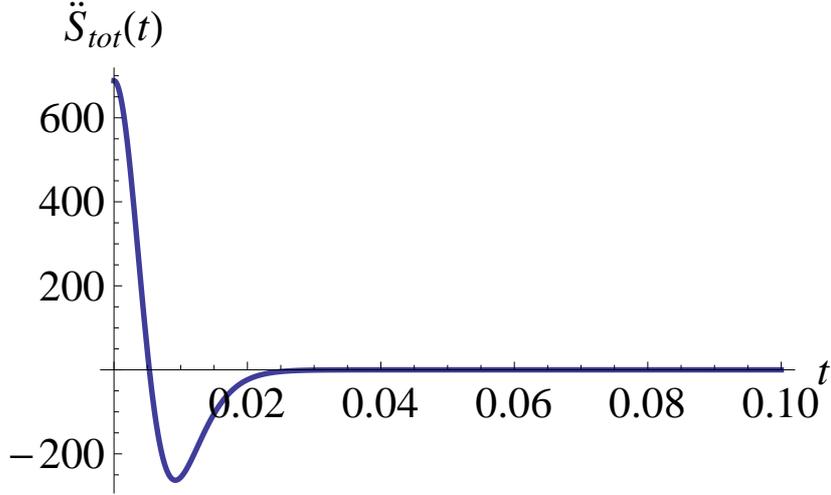}
\caption{The diagram presents the evolution of $\ddot S_\text{tot}$ in the case when $\rho_\text{de}=\Lambda_\text{bare}+\frac{\alpha_{2}}{t^2}$, where $\alpha_\text{2}<0$. The cosmological time $t$ and $\ddot S_\text{tot}$ are expressed in $\frac{\text{s}\ \text{Mpc}}{\text{km}}$ and $\frac{\text{s}^2\ \text{Mpc}^2}{\text{km}^2}$, respectively. Note that $\ddot S_\text{tot}$ approaches asymptotically zero as $t$ tends to the infinity. $\ddot S_\text{tot}$ approaches zero from below for the late times. Therefore, the convexity condition is satisfied and in consequence, the entropy will never be unbounded. This fact has a simple interpretation that the Universe tends towards a state of the thermodynamic equilibrium.}
\label{fig3}
\end{figure}

We can obtain the form of the variation of the second derivative of entropy $\ddot S_\text{tot}$ from eq.~(\ref{entropy}). Then we get
\begin{equation}
\begin{split}
\ddot S_\text{tot} &= \frac{18\pi^2}{\tilde r_A^2}H^2\left[(1+w_\text{de})\rho_\text{de}+(1+w_\text{m})\rho_\text{m}\right]
\Big[\Big((1+w_\text{de})\rho_\text{de}+(1+w_\text{m})\rho_\text{m}\Big) \times \\
&\times\Big((1+5w_\text{de}-4w_\text{m})\rho_\text{de} +(1+w_\text{m})\rho_\text{m}\Big)+4(w_\text{de}-w_\text{m})H\dot\rho_\text{de}\Big].
\end{split}
\end{equation}
The evolution of $\ddot S_\text{tot}$ for the case $\rho_\text{de}=\Lambda_\text{bare}+\frac{\alpha_{2}}{t^2}$ is presented in figure~\ref{fig3}. One can observe that the second derivative of entropy approaches zero from below in the late time evolution of the Universe. Hence the convexity condition is satisfied in the final states of evolution and as a result entropy will never grow unbounded in this case.

\section{The evolution of the temperature in models with a quantum decay of the vacuum}

To study the evolution of temperature of dark matter, we assume ideal gas as a model of dark matter \cite{Armendariz-Picon:2013jej, Delort:2018}. In this case, we cannot treat cold dark matter as dust and in result we assume that $w_\text{m}\neq 0$ and $w_\text{m}$ is not constant. For the perfect fluid, energy momentum tensor $T^{\alpha\beta}$ is given by
\begin{equation}
T^{\alpha\beta}=(\rho+p)u^\alpha u^\beta-pg^{\alpha\beta},\label{tensor}
\end{equation}
where $g^{\alpha\beta}$ is the metric and $u^\alpha$ is the four velocity.
From equation (\ref{tensor}), we can obtain the energy conservation law $\left(u_{\alpha}T^{\alpha\beta}_{;\beta}=0\right)$
\begin{equation}
\dot\rho+3H(\rho+p)=0.\label{conservation}
\end{equation}
Here, $_{;\beta}\equiv\nabla_\beta$ denotes divergence.

Let $N^\alpha$ be the particle current, which is defined by
\begin{equation}
N^\alpha=nu^\alpha,
\end{equation}
where $n$ is the particle number density of the fluid component. If the number of particles is not conserved then
\begin{equation}
N^\alpha_{;\alpha}=Q
\end{equation}
or
\begin{equation}
\dot n+3Hn=Q,\label{particles}
\end{equation}
where $Q$ is the interacting term defined by
\begin{equation}
Q=-\dot\Lambda.\label{interaction}
\end{equation}
When $Q>0$, then the number of particles increases with the time. When $Q<0$, then the number of particles decreases with the time.

Let $w_\text{de}=-1$. Because we assume in our model that matter is described by ideal gas, the equation of state for matter is given by $p_\text{m}(t)=w_\text{m}(t)\rho_\text{m}(t)=n(t)kT(t)$, where $n$ is number of particles per unit volume, and $k$ is the Boltzmann constant. Because for non-relativistic dark matter \cite{Armendariz-Picon:2013jej, Delort:2018}
\begin{equation}
\rho_\text{m}(t)=mn(t)+n(t)kT(t)
\end{equation}
or
\begin{equation}
n(t)=\frac{\rho_\text{m}(t)}{m+kT(t)},\label{nn}
\end{equation}
we obtain
\begin{equation}
\dot\rho_\text{m}(t)=\dot n(t)\left(m+kT(t)\right)+ n(t)k\dot T(t),\label{state}
\end{equation}
where $m$ is mass of dark matter particle.
The energy conservation equation (\ref{conservation}) can be rewritten as
\begin{equation}
\dot\rho_\text{m}(t)=-3H(t)(\rho_\text{m}(t)+p_\text{m}(t))-\dot\Lambda(t)\label{continuous}
\end{equation}
or
\begin{equation}
\dot n(t)\left(m+kT(t)\right)+ n(t)k\dot T(t)=-3H(t)\left(\rho_\text{m}(t)+n(t)kT(t)\right)-\dot\Lambda(t).\label{continuous2}
\end{equation}

From eqs.~(\ref{particles}) and (\ref{interaction}), we obtain that
\begin{equation}
-\dot\Lambda(t)=\dot n(t)+3H(t)n(t).\label{continuous3}
\end{equation}

The above equation can be rewritten as
\begin{equation}
\dot N(t)=-\dot\Lambda(t)a(t)^3,
\end{equation}
where $N=na^{3}$ is the number of particles.

From eqs.~(\ref{nn}), (\ref{continuous2}) and (\ref{continuous3}), we obtain
\begin{equation}
\dot T(t)=-3H(t)T(t)+\frac{\left(m+k T(t)-1\right)\left(m+k T(t)\right)\dot\Lambda(t)}{k\rho_\text{m}(t)}.\label{temperature}
\end{equation}
In the case $\Lambda(t)=\Lambda_\text{bare}+\frac{\alpha_{2}}{t^2}+\frac{\alpha_{4}}{t^4}$, eq.~(\ref{temperature}) has the following form
\begin{equation}
\dot T(t)=-3H(t)T(t)-\frac{\left(m+k T(t)-1\right)\left(m+k T(t)\right)\left(\frac{2\alpha_2}{t^3}+\frac{4\alpha_4}{t^5}\right)}{k\rho_\text{m}(t)}.
\end{equation}
When we neglect $\Lambda_\text{bare}$ and use respectively eq.~(\ref{bessel1}) and eq.~(\ref{bessel2}), then the above equation gets respectively the following form for $\alpha_4>0$
\begin{equation}
\begin{split}
\dot T(t)=-3\left(\frac{1-\sqrt{1+3(w_\text{m}(t)+1)^2 \alpha_2}}{3(w_\text{m}(t)+1)t} -\frac{\sqrt{\alpha_4}I_{n+1}\left(\frac{(w_\text{m}(t)+1)\sqrt{3\alpha_4}}{2t}\right)}{\sqrt{3}t^2 I_{n}\left(\frac{(w_\text{m}(t)+1)\sqrt{3\alpha_4}}{2t}\right)}\right)T(t)\\-\frac{\left(m+k T(t)-1\right)\left(m+k T(t)\right)\left(\frac{2\alpha_2}{t^3}+\frac{4\alpha_4}{t^5}\right)}{k\left(3\left(\frac{1-\sqrt{1+3(w_\text{m}(t)+1)^2 \alpha_2}}{3(w_\text{m}(t)+1)t} -\frac{\sqrt{\alpha_4}I_{n+1}\left(\frac{(w_\text{m}(t)+1)\sqrt{3\alpha_4}}{2t}\right)}{\sqrt{3}t^2 I_{n}\left(\frac{(w_\text{m}(t)+1)\sqrt{3\alpha_4}}{2t}\right)}\right)^2-\frac{\alpha_{2}}{t^2}-\frac{\alpha_{4}}{t^4}\right)},
\end{split}
\end{equation}
and for $\alpha_4<0$
\begin{equation}
\begin{split}
\dot T(t)=-3\left(\frac{1-\sqrt{1+3(w_\text{m}(t)+1)^2 \alpha_2}}{3(w_\text{m}(t)+1)t} + \frac{\sqrt{-\alpha_4}J_{n+1}\left(\frac{(w_\text{m}(t)+1)\sqrt{-3\alpha_4}}{2t}\right)}{\sqrt{3}t^2 J_{n}\left(\frac{(w_\text{m}(t)+1)\sqrt{-3\alpha_4}}{2t}\right)}\right)T(t)\\-\frac{\left(m+k T(t)-1\right)\left(m+k T(t)\right)\left(\frac{2\alpha_2}{t^3}+\frac{4\alpha_4}{t^5}\right)}{k\left(3\left(\frac{1-\sqrt{1+3(w_\text{m}(t)+1)^2 \alpha_2}}{3(w_\text{m}(t)+1)t} + \frac{\sqrt{-\alpha_4}J_{n+1}\left(\frac{(w_\text{m}(t)+1)\sqrt{-3\alpha_4}}{2t}\right)}{\sqrt{3}t^2 J_{n}\left(\frac{(w_\text{m}(t)+1)\sqrt{-3\alpha_4}}{2t}\right)}\right)^2-\frac{\alpha_{2}}{t^2}-\frac{\alpha_{4}}{t^4}\right)},
\end{split}
\end{equation}
where $w_\text{m}(t)=\frac{kT(t)}{m+kT(t)}$.

\begin{figure}
\centering
\includegraphics[width=0.7\linewidth]{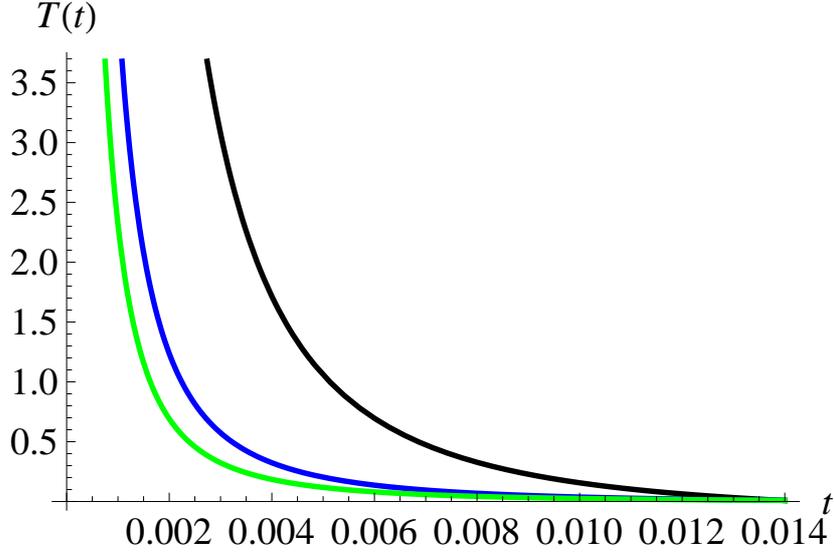}
\caption{The diagram presents the evolution of the temperature $T(t)$ for the model with $\rho_\text{de}=\Lambda_\text{bare}+\frac{\alpha_{2}}{t^2}$ for different values of the mass of particle of dark matter for the case $m=100$ keV (black curve), $m=10$ keV (blue curve), and $m=1$ keV (green curve). Here, $\alpha_2=-10^{-4}\frac{\text{s}\ \text{Mpc}}{\text{km}}$. The cosmological time $t$ is expressed in $\frac{\text{s}\ \text{Mpc}}{\text{km}}$ (here, the present age of the Universe ${\cal T}_{0}=0.014\frac{\text{s}\ \text{Mpc}}{\text{km}}$) and $T(t)$ is expressed in Kelvin (the present value is assumed as 0.01 K).}
\label{fig5}
\end{figure}

\begin{figure}
\centering
\includegraphics[width=0.7\linewidth]{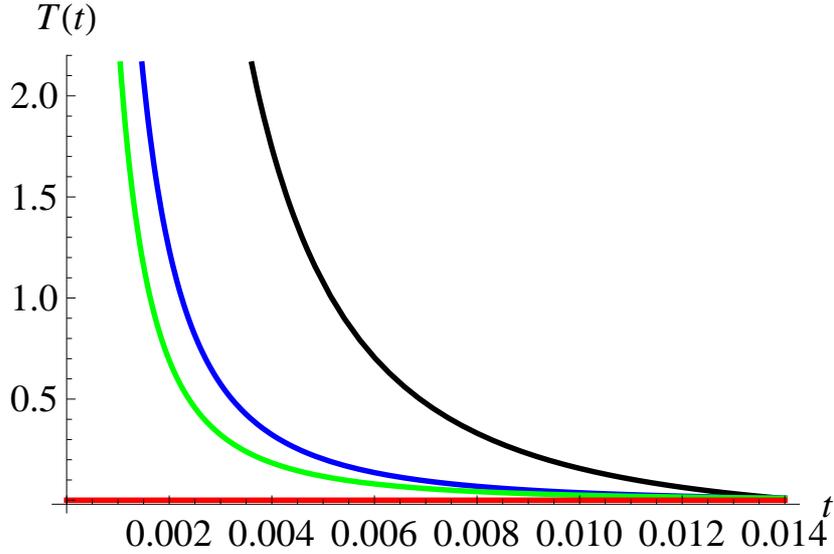}
\caption{The diagram presents the evolution of the temperature $T(t)$ for the model with $\rho_\text{de}=\Lambda_\text{bare}+\frac{\alpha_{2}}{t^2}$ for the case $\alpha_2=-10^{-2}\frac{\text{s}\ \text{Mpc}}{\text{km}}$ (black curve), $\alpha_2=-10^{-3}\frac{\text{s}\ \text{Mpc}}{\text{km}}$ (blue curve), $\alpha_2=-10^{-4}\frac{\text{s}\ \text{Mpc}}{\text{km}}$ (green curve), and $\alpha_2=0$ (red curve). Here, the mass of particles of dark matter $m$ is equal to 1 keV. The cosmological time $t$ is expressed in $\frac{\text{s}\ \text{Mpc}}{\text{km}}$ (here, the present age of the Universe ${\cal T}_{0}=0.014\frac{\text{s}\ \text{Mpc}}{\text{km}}$) and $T(t)$ is expressed in Kelvin (the present value is assumed as 0.01 K).}
\label{fig7}
\end{figure}

Figures~\ref{fig5} and \ref{fig7} present the evolution of $T(t)$ for $\Lambda(t)=\Lambda_\text{bare}+\frac{\alpha_{2}}{t^2}+\frac{\alpha_{4}}{t^4}$. In Fig.~\ref{fig5}, we can observe a growth of temperature caused by quantum effects of decaying of dark energy in the early time universe.

In the adiabatic photon creation case \cite{Lima:1995kd}, we have an assumption
\begin{equation}
p^{1-\gamma}T^\gamma=\text{const}
\end{equation}
or in the equivalent form
\begin{equation}
n^{1-\gamma}T=\text{const}\label{particles3}
\end{equation}
or
\begin{equation}
(Na^{-3})^{1-\gamma}T=\text{const}.\label{particles2}
\end{equation}
Because we assume that $p=w_\text{m}\rho_\text{m}=(\gamma-1)\rho_\text{m}$ \cite{Lima:1995kd}, we obtain that
\begin{equation}
\rho_\text{m}=dT^\frac{\gamma}{\gamma-1},\label{adiabatic}
\end{equation}
where $d=\text{const}$, $w_\text{m}+1=\gamma=\text{const}$, and $\gamma\neq 1$.

After using eq.~(\ref{adiabatic}) in the Friedmann equation (\ref{friedmann}) we get that
\begin{equation}
3H(t)^2=dT(t)^\frac{\gamma}{\gamma-1}+\Lambda(t).
\end{equation}
In consequence, we can easily obtain that
\begin{equation}
T(t)=\left(\frac{3H(t)^2-\Lambda(t)}{d}\right)^{\frac{\gamma-1}{\gamma}}\label{adiabatic3}
\end{equation}
or after respectively using eq.~(\ref{particles2}) and eq.~(\ref{particles3})
\begin{equation}
n=\left(\frac{3H(t)^2-\Lambda(t)}{b}\right)^{\frac{1}{\gamma}}
\end{equation}
and
\begin{equation}
N=a^3\left(\frac{3H(t)^2-\Lambda(t)}{b}\right)^{\frac{1}{\gamma}},
\end{equation}
where $b=\text{const}$.
In the case of $\Lambda(t)=\Lambda_\text{bare}+\frac{\alpha_{2}}{t^2}+\frac{\alpha_{4}}{t^4}$, eq.~(\ref{adiabatic3}) has the following form
\begin{equation}
T(t)=\left(\frac{3H(t)^2-\Lambda_\text{bare}-\frac{\alpha_{2}}{t^2}-\frac{\alpha_{4}}{t^4}}{d}\right)^{\frac{\gamma-1}{\gamma}}.\label{adiabatic2}
\end{equation}

Because $\rho_\text{r}=\rho_\text{r,0}a^{-4+\lambda(t)}$, for the radiation case ($\gamma=4/3$), we can obtain that
\begin{equation}
T=T_0 a^{\lambda/4-1},
\end{equation}
where $T_0$ is the present value of temperature of the Universe.

When we neglect $\Lambda_\text{bare}$ term and use eq.~(\ref{bessel1}) and eq.~(\ref{bessel2}), respectively, eq.~(\ref{adiabatic2}) gets the following form for $\alpha_4>0$
\begin{equation}
T(t) = \Bigg\{ \frac{1}{d} \Bigg[ 3\Bigg(\frac{1-\sqrt{1+3(w_\text{m}+1)^2 \alpha_2}}{3(w_\text{m}+1)t} -\frac{\sqrt{\alpha_4}I_{n+1}\big(\frac{(w_\text{m}+1)\sqrt{3\alpha_4}}{2t}\big)}{\sqrt{3}t^2 I_{n}\big(\frac{(w_\text{m}+1)\sqrt{3\alpha_4}}{2t}\big)}\Bigg)^2-\frac{\alpha_{2}}{t^2}-\frac{\alpha_{4}}{t^4}\Bigg]\Bigg\}^{\frac{\gamma-1}{\gamma}}
\end{equation}
and for $\alpha_4<0$
\begin{equation}
T(t) = \Bigg\{ \frac{1}{d} \Bigg[ 3\Bigg(\frac{1-\sqrt{1+3(w_\text{m}+1)^2 \alpha_2}}{3(w_\text{m}+1)t} +\frac{\sqrt{-\alpha_4}J_{n+1}\big(\frac{(w_\text{m}+1)\sqrt{-3\alpha_4}}{2t}\big)}{\sqrt{3}t^2 J_{n}\big(\frac{(w_\text{m}+1)\sqrt{-3\alpha_4}}{2t}\big)}\Bigg)^2-\frac{\alpha_{2}}{t^2}-\frac{\alpha_{4}}{t^4}\Bigg]\Bigg\}^{\frac{\gamma-1}{\gamma}}.
\end{equation}

Our general conclusion arising from the thermodynamic analysis is that the model seems to be very natural. The dynamics of the Universe is irreversible when full quantum effects of the decaying vacuum are taken into account. Of course, this irreversibility has a thermodynamic interpretation as far as the evolution of the Universe is concerned. In thermodynamics, the area of the cosmological horizon is interpreted as a black hole Hawking entropy. In the model under consideration, the entropy growth is an irreversible process and entropy grows to the state of a thermal equilibrium. This process is independent on details of quantum corrections.

\section{Conclusions}

From our investigation of cosmological implications of the quantum decay of metastable dark energy, we got the following results.

Firstly, the cosmological models with the running cosmological parameter are an extension of the $\Lambda$CDM model. The new ingredient in the comparison with the standard cosmological model ($\Lambda$CDM model) is the interaction between the matter and dark energy. In result, the canonical scaling law $\rho_\text{m}\propto a^{-3}$ is modified. Since $\Lambda(t)$ is decaying ($\frac{d\Lambda}{dt}<0$) density of matter in the comoving volume $\propto a^{3}$ increases with time. The leading term at late times of the evolution of the Universe in $\Lambda(t)$ is of the order $1/t^2$.

The analogous conclusion was formulated by Freese et al. \cite{Freese:1986dd} in the context of the decaying vacuum. They considered the case when the particle number is not conserved $\frac{d}{dt}\left(\rho_\text{m}a^3\right)\neq 0$ during the cosmic evolution which implies the continuous creation of massive particles. If we denote by $\Delta n_B$ the net baryon (or antibaryon) number density created by the vacuum during the matter dominated epoch then $\Delta n_B (t)\propto t^{-2}$ and $\rho_v \propto t^{-2}$ (see formulas (31) and (30b) in ref.~\cite{Freese:1986dd}).

The parameterization of the metastable dark energy in terms of the power series with respect to $1/t$ is symmetric (time reflection $t \to - t$) because it contains only even terms. For the present time of the cosmic evolution only the leading term $1/t^2$ prevails. This term is a residue from earlier epochs. If we express this term in terms of size of horizon $L$ than this residual metastable dark energy is proportional $L^{-2}$. Note that this term assumes the form of holographic dark energy. In the holographic universe dark energy is quantum zero-point energy density caused by distance cut-off, so the total energy inside of the sphere with radius $L$ is equal to mass of a black hole of the same size $\rho_{\Lambda} = 3 c^2 L^{-2}$ \cite{Li:2004rb}. In some sense, if we take full parameterization of metastable dark energy in a Taylor series, this formula can be interpreted as the generalized model of holographic dark energy, replacing $t$ by the size of horizon $L$.

Secondly, the generalized second law of thermodynamics required that, the total entropy, due to matter, quantum corrections and horizon together, must always increase, i.e., $\frac{dS}{dt}\geq 0$. This condition is always satisfied in the model under the consideration. We obtained this condition in the same form like in the case without quantum corrections.

On the other hand the function $\frac{d^2S}{dt^2}$ approaches zero from below during the late time evolution of the Universe (see figure~\ref{fig3}). Therefore, the upper convexity condition $\frac{d^2S}{dt^2}<0$ is satisfied at final stages of the evolution of the Universe. This fact has a simple interpretation that the Universe goes towards a state of the thermodynamic stable equilibrium.

For a better interpretation \cite{Alcaniz:2005dg} of the vacuum decay scenario we considered the temperature evolution of the Universe with respect to the cosmological time. We considered the photon creation process in the adiabatic form. As we have seen, in the ``adiabatic'' case \cite{Lima:1995kd}, the temperature $T$ satisfies (\ref{adiabatic3}) dependence on some details of the vacuum decay. For the early Universe, we found that the quantum effects modified the evolution of the temperature (see figures~\ref{fig5} and \ref{fig7}). It is found that the vacuum decaying into photon obeys the generalized scaling relation $\rho_\text{r}=\rho_\text{r,0}a^{-4+\lambda(t)}$ and its temperature scales as $T(t)=T_0 a(t)^{\lambda/4-1}$. We also argued that in the context of the thermodynamic consideration the basic parameter quantifying the vacuum decay rate must be positive.

We also found that the adiabatic condition $\frac{|\dot \rho_\text{de}|}{|\rho_\text{de}|}\gg\frac{\dot a}{a}$ is violated when $\beta>10^{7}$ or $\rho_\text{de}^0>10^{93}\Lambda_\text{bare}$.

To complete the discussion, let us recall that the decay rate of the system in the metastable vacuum state was discussed and analyzed in \cite{Szydlowski:2017wlv}, where it was represented graphically as a function of time (see eqs~(10), (11) and fig.~2 in ref.~\cite{Szydlowski:2017wlv}).

From our consideration of the decaying process of metastable dark energy one can also conclude that the early Universe can offer some possibilities of testing of quantum mechanics. This fundamental theory can be probed in a standard way in labs but also in different regimes established by the early Universe. It seems to be a natural way of testing the limits of quantum mechanics.

\providecommand{\href}[2]{#2}\begingroup\raggedright\endgroup
\end{document}